\documentclass[twocolumn,amsmath,amssymb,nofootinbib,eqsecnum,tighten,floatfix,prb]{revtex4-1}

\usepackage{graphicx}
\usepackage{dcolumn} 
\usepackage{bm, bbm}      
\usepackage{curves}
\usepackage{epic}
\usepackage{wasysym}
\usepackage{epsf, times}
\usepackage{subfigure}
\usepackage{eufrak}
\usepackage{amsthm}

\def\bb{{\bf b}}
\def\F{\mathcal{F}}
\def\ii{\mathbf{i}}
\def\jj{\mathbf{j}}
\def\kk{\mathbf{k}}
\def\s{\sigma}

\def\xx{{\bf x}} \def\yy{{\bf y}} \def\zz{{\bf z}}

\def\N{\mathcal{N}}

\def\C{\mathcal{C}}
\def\Cp{{\mathcal{C}^{\prime}}}
\def\D{\mathcal{D}}
\def\O{\mathcal{O}}
\def\openone{\leavevmode\hbox{\small1\kern-4.2pt\normalsize1}}

\newcommand{\beq}{\begin{equation}}
\newcommand{\eeq}{\end{equation}}
\newcommand{\bea}{\begin{eqnarray}}
\newcommand{\eea}{\end{eqnarray}}
\newcommand{\llangle}{\left\langle}
\newcommand{\rrangle}{\right\rangle}
\newcommand{\llvert}{\left\vert}
\newcommand{\rrvert}{\right\vert}

\usepackage[dvips]{color}

\begin{document}


\title{
Quantum mechanical and information theoretic view 
on classical glass transitions
}

\author{
Claudio Castelnovo$^1$, 
Claudio Chamon$^2$, 
and 
David Sherrington$^1$
       }
\affiliation{
$^1$ 
Rudolf Peierls Centre for Theoretical Physics, 
University of Oxford, Oxford, OX1 3NP, UK
\\ 
$^2$ 
Physics Department, Boston University, Boston, MA 02215, USA
            }

\date{\today}

\begin{abstract}
Using the mapping of the Fokker-Planck description of classical
stochastic dynamics onto a quantum Hamiltonian, we argue that a
dynamical glass transition in the former must have a precise
definition in terms of a quantum phase transition in the latter. At
the dynamical level, the transition corresponds to a collapse of the
excitation spectrum at a critical point. At the static level, the
transition affects the ground state wavefunction: while in some cases
it could be picked up by the expectation value of a local operator, in
others the order may be non-local, and impossible to be determined
with any local probe. Here we instead propose to use concepts from
quantum information theory that are not centered around local order
parameters, such as fidelity and entanglement measures. 
We show that for systems
derived from the mapping of classical stochastic dynamics, 
singularities in the fidelity susceptibility translate directly into 
singularities in the heat capacity of the classical system. 
In classical glassy systems with an extensive number of metastable 
states, we find that the prefactor of the area law term in the 
entanglement entropy jumps across the transition. 
We also discuss how entanglement measures can be used to detect a growing 
correlation length that diverges at the transition. 
Finally, we illustrate how static order can be hidden in systems with a 
macroscopically large number of degenerate equilibrium states 
by constructing a three dimensional lattice gauge model with
only short-range interactions but with a finite temperature continuous phase
transition into a massively degenerate phase. 
\end{abstract}

\maketitle
%
%

\section{\label{sec: intro}
Introduction
        } 
Order and disorder are fundamental concepts in understanding phases
of matter. A classification of states of matter is possible according
to the symmetries that are broken in the condensed
state,~\cite{Anderson1972} which can be detected through the
non-vanishing expectation value of a local observable: the order
parameter. Crystalline order for instance is characterized by the
breaking of full translational symmetry into a regular periodic
lattice. Glasses, however, do not have a (spatially) local order parameter 
that can be defined from a single snapshot of the 
microscopic configuration of the physical degrees of freedom, say, the
position of the constituent atoms. 
[There are order parameters in replica type theories, 
but they require correlations across replicas.~\cite{replica_theory}] 
Hence, glass order does not fall within the Landau paradigm of classification 
via symmetry and local order parameters. 

It has been argued for many years that the glass transition might have
a purely dynamical origin, and that it is not accompanied by a
thermodynamic (static) transition. 
This scenario prompts then a fundamental question: Is it possible
to define accurately a {\it dynamic} phase transition, and 
to understand why {\it static} order 
appears to be absent? 

In this paper, we show that a positive answer to this question can be
found by exploiting the mapping between the stochastic dynamics of a
classical system and a quantum mechanical system at zero
temperature. 
The mapping is based on a well-known symmetrization of the classical 
transition matrix,~\cite{symmetric_W_refs}, provided it satisfies detailed 
balance, and on the promotion of the classical configurations to 
orthonormal basis states of a Hilbert space. 
This yields a correspondence between the classical system and a quantum 
Hamiltonian \emph{in the preferred configuration basis}, with two fundamental 
properties: 
(i) all quantum ground state correlators of diagonal operators in the 
preferred basis are equivalent to classical correlation functions; 
and 
(ii) the energy spectrum of the Hamiltonian equals the spectrum of relaxation 
rates in the stochastic classical system. 
In particular, the longest equilibration time scale in the classical 
system is inversely proportional to the splitting between the 
ground state and the first excited state of the quantum Hamiltonian. 
(See Ref.~\onlinecite{Castelnovo2005} for a constructive formulation of 
this correspondence.) 
In essence, the mapping merges a $d$-dimensional classical system with 
stochastic dynamics (i.e., with an additional `time dimension') into a 
$d$-dimensional quantum system. 
In the quantum mechanical language, concepts in quantum phase transitions and 
quantum information can be used to formulate precisely \emph{what} a 
dynamical transition \emph{is} in the original classical system, and provide 
new avenues to study these transitions. 

In conventional approaches to dynamical transitions, one often relies 
on time dependent concepts, such as metastable states around free energy 
minima. 
These concepts require the ability to distinguish between short and long 
relaxation time scales, that characterise for instance the relaxation of 
the system within a basin, and transitions between basins, respectively. 
This in turn introduces an intrinsic difficulty in comparing the behaviour 
of finite size systems with respect to the thermodynamic limit. 

Note however that the quantum Hamiltonian contains \emph{all} the information 
about the original classical stochastic process, and, as such, one is not 
required to look at time dependent quantities. 
In recent years, a view of stochastic dynamics as ``a kind of thermodynamics 
in space-time''~\cite{Kurchan2009} was developed using large deviation 
functions.~\cite{large_dev_refs,large_dev_firstorder_refs} 
In this theory, one weighs trajectories by a ``Boltzmann measure'' that 
couples to their `activity', thus favouring slow vs fast relaxing modes 
in the system, depending on the sign of the coupling constant. 
In the quantum mechanical language, the weights amount to perturbing terms 
in the Hamiltonian which break detailed balance, and transitions are studied 
as a function of the coupling constant. 
In a sense, large deviation functions can be seen as a way to probe the 
physics of the quantum system by looking at how it responds to external
perturbations. 
Relations between large deviations in classical systems and 
quantum critical points have been discussed recently in 
Ref.~\onlinecite{Jack2009}, and further elaborated in 
Ref.~\onlinecite{Elmatad2010}.

In this paper, we take a different point of view. 
By committing to the study of the 
quantum mechanical system per se, we show that one can detect, 
characterise and possibly classify glass transitions as peculiar 
\emph{static} quantum phase transitions. 

If the classical system undergoes 
a dynamical transition, the gap must close in the associated quantum 
model, leading to a degeneracy that persists throughout the incipient 
phase. Such gapped-to-gapless transition signals therefore a 
(thermodynamic) quantum phase transition that provides the precise 
definition of a dynamical transition in the classical system. 
(Generically, soft modes can be present also in the fast-relaxing, 
high-temperature phase of the system. In Sec.~\ref{sec: q_EA} 
we discuss how these soft modes ought to be treated separately from the 
spectral collapse responsible for a glass transition.) 
The appearance of a non-vanishing Edwards-Anderson order 
parameter~\cite{Edwards1975} across the transition translates directly into 
a divergent static local susceptibility in the quantum system. 

One of the great conceptual difficulties in defining a dynamical 
transition is the fact that there seems to be no local order parameter 
differentiating the phases on the two sides of the transition. A 
glass looks locally and instantaneously like a liquid, much as a spin glass 
looks like a paramagnet. 
Here we argue that using the quantum mechanical mapping one can do away with 
local order parameters in the classical configuration basis, 
and one can effectively investigate static signatures of a dynamical glass 
transition using off-diagonal operators and non-local measures, which are more 
natural in the quantum-equivalent formulation of the stochastic problem. 

A quantum phase transition is signaled by changes in the
ground state wavefunction. For example, if there is a local order
parameter, the transition can be picked up by the expectation value of
local operators. These operators need not be diagonal in the preferred 
basis given by the classical configurations. For example, in a spin
system where the classical basis is the $z$-direction, ordering in the
$x$-direction in the quantum system amounts to a `dynamical' order in
its classical counterpart. 
The quantum mechanical language gives immediate access to operators written 
in a form that is not diagonal in the preferred basis, and therefore it 
broadens the field of search for local order parameters that might capture 
a glass transition. 

Moreover, quantum phase transitions can be studied using measures that 
do not rely on the existence (and therefore, on any \textit{a priori} 
knowledge) of an order parameter. 
For instance, concepts from quantum information theory, such as fidelity 
susceptibility~\cite{fidelity_refs,fidelity_susc_refs} and 
entanglement entropy~\cite{Amico-etal} have proven useful in detecting and 
characterising transitions into exotic phases with non-local order parameters 
(e.g., topologically ordered 
phases~\cite{Castelnovo2008a,Abasto2008,QPTbookChapter}). 
Motivated by the difficulty in finding any static order accompanying
dynamical transitions, we propose to employ such measures 
in this context. 

The fidelity is constructed from the overlap of the 
wavefunction at two infinitesimally close values of the coupling 
constant that drives the transition.~\cite{fidelity_refs} 
It can be expressed straightforwardly 
in the basis where the classical configurations are defined. 
Here, we show that, for quantum Hamiltonians derived from the stochastic 
dynamics of a classical system, the quantum fidelity susceptibility is 
proportional to the heat capacity of the classical system. A 
singularity in the fidelity susceptibility at the phase transition 
translates therefore into a singularity in the heat capacity of the 
classical system. The latter is a feature that is indeed present in 
several theoretical and experimental examples of glass 
transitions,~\cite{glass_expm} but upon which there is so far no consensus 
in the literature. 

Another tool to study quantum phase transitions, even exotic ones
without a local order parameter, is the von Neumann entanglement
entropy. Scaling properties of the von Neumann entropy as a function
of the size of the subsystem, for instance, can reveal a correlation
length that diverges at the phase transition in topologically ordered 
systems. It is thus interesting
to examine what entanglement entails for the quantum system derived 
by mapping the dynamics of a glass. Here we discuss how the 
von Neumann entanglement entropy of a bipartition of the systems is 
sensitive to the properties of the two classical subsystems, and to 
the way they influence one another. For instance, in the case of 
glassy phases with an extensive number of non-relaxing modes (in the 
thermodynamic limit), we show that the entanglement entropy obeys the 
(conventional) area law both above and below the transition, but with a 
prefactor that jumps across the phase transition. 
As we discuss, one can also use it to detect a diverging correlation length 
at the dynamical transition. 

The paper is organised as follows. 
We begin by re-deriving typical (time dependent) quantities used 
to characterise glass transitions in the quantum mechanical language. 
For example, the Edwards-Anderson order parameter,~\cite{Edwards1975} 
defined as the appropriate limit of a connected autocorrelation function, 
can be straightforwardly written in quantum mechanical language 
to show that a phase with a non-vanishing $q_{\rm EA}$ is in fact 
characterized by a divergent static susceptibility. 
The susceptibility is computed with respect to a perturbation to the 
mapped quantum Hamiltonian that cannot be written as an equivalent 
perturbation of the original classical system. This perturbation is 
``quantum'' in nature, akin to the one used in large deviation theory. 
[The familiar reader might want to move quickly through Sec.~\ref{sec: method} 
and Sec.~\ref{sec: q_EA}.] 

We then enter the discussion of \emph{static} quantum mechanical measures 
that can be used to investigate glass transitions. 
In particular, we focus on quantities that probe phase transitions 
without resorting to local order parameters: the quantum information 
theoretic concepts of fidelity susceptibility 
(Sec.~\ref{sec: fidelity}) and entanglement entropy 
(Sec.~\ref{sec: entanglement}). 

Finally, in Sec.~\ref{sec: KCM example} we construct an explicit example where 
we can analytically show that there is a static phase transition accompanying 
a dynamical one, in a system where the order is hidden. This example, a 
generalization of the three-dimensional gonihedric spin model, has a 
finite temperature transition into a phase with a macroscopic number 
of equilibrium states. If endowed with gaugeable disorder, it is 
impossible to select a single minimum-energy configuration by applying a 
local field. 

In order to ensure a clear flow through the central results, 
several parts of the paper have been moved to the appendix. 
In Appendix~\ref{app: collapsing states}, we discuss the nature of the 
collapsing states at the transition. 
In Appendix~\ref{app: two-body examples}, 
as an example of the mapping of classical stochastic dynamics to quantum 
Hamiltonians, we carry out explicitly the construction for the 
Sherrington-Kirkpatrick model, and we explain the connection between the 
Parisi equilibrium minima and the degenerate quantum states. 
In Appendix~\ref{app: variational excited states}, 
we show how one can construct variational wavefunctions for the case of 
symmetry-broken states, and how ergodicity breaking can be related to 
their vanishing variational energies. 
In Appendix~\ref{app: gonihedric model}, 
by writing the gonihedric model in terms of dual variables, 
we construct an example of a (toric-code-like~\cite{Kitaev2003}) 
quantum system with a macroscopic topological ground state degeneracy. 
%
%

\section{\label{sec: method}
Review of the mapping of stochastic processes to quantum Hamiltonians
        }
The fact that a transition matrix $W$ in a Markov process (obeying detailed 
balance) can be made into a symmetric matrix by means of a similarity 
transformation has long been exploited to either find the eigenvalues and 
eigenvectors of $W$,~\cite{symmetric_W_refs} or to gain insight on quantum 
Hamiltonians by turning them into classical dynamical 
systems.~\cite{RK_like_refs} 

In Ref.~\onlinecite{Castelnovo2005} it was shown that this correspondence is 
indeed one-to-one onto a class of quantum Hamiltonian representations, dubbed 
Stochastic Matrix Form (SMF) decompositions, that essentially encompasses 
all quantum systems devoid of the sign problem. 

In this mapping, the temperature $T=1/\beta$ of the classical system enters 
the quantum Hamiltonian as a coupling constant. 
In particular, equilibrium correlators in the classical system at temperature 
$T$ map onto equal time ground state (GS) correlators in the quantum system, 
and the Markov relaxation rates become the spectrum of the quantum 
Hamiltonian.

The study of eigenvalue degeneracy in dynamical systems, suggested by Kac 
amongst others, has long been exploited in trying to characterise the 
appearance of long-lived metastable states, and glassiness 
(see for instance Refs.~\onlinecite{Gaveau1998},\onlinecite{Biroli2001}, 
and references therein). 

Consider a system whose configurations are labeled by $\C$, 
and where the probability $P_\C(t)$ of the system being in configuration 
$\C$ at time $t$ is governed by the Markov equation 
\beq
\frac{d}{dt} P_\C(t) 
= 
\sum_{\Cp} W_{\C,\Cp} P_{\Cp}(t) 
, 
\label{eq:Markov-eq}
\eeq
where the transition matrix $W_{\C,\Cp}$ satisfies
probability conservation
$
W_{\C,\C} = - \sum_{\Cp \neq \C} W_{\C,\Cp} 
, 
$
and detailed balance
$
W_{\C,\Cp} \, e^{- E_\C / T}
= 
W_{\Cp,\C} \, e^{- E_{\Cp} / T} 
. 
$

It is convenient to introduce the following vector notation. We can consider 
$P_\C(t)$ as being the real component of a vector $\llvert P(t) \rrangle$ in 
the span of the orthonormal basis $\{ \llvert \C \rrangle, \; \forall\,\C \}$, 
%
%
$
P_\C(t) = \llangle \C \vert P(t) \rrangle 
$. 
%
%
Eq.~\eqref{eq:Markov-eq} becomes then 
\beq
\frac{d}{dt} \llvert P(t) \rrangle 
= 
\hat{W} \llvert P(t) \rrangle 
, 
\label{eq: Markov-eq-vec} 
\eeq
which admits solutions in the form 
\beq
\llvert P(t) \rrangle 
= 
\exp \left( \hat{W} t \right) 
\llvert P(0) \rrangle 
, 
\eeq
where we used the fact that $\hat{W}$ does not depend explicitly on $t$. 
Note that $W_{\C,\Cp}$ is real but not necessarily symmetric, hence 
$\hat{W}$ needs not be Hermitian, and one has to distinguish between 
left and right eigenvectors. 

The right eigenvectors are given by 
\beq
\hat{W} \llvert \psi_R^{(n)} \rrangle 
= 
- \varepsilon_n \, \llvert \psi_R^{(n)} \rrangle 
, 
\eeq
where $\varepsilon_0 = 0$, $\varepsilon_n > 0$ for $n \neq 0$ 
(as ensured by Perron-Frobenius theorem for any irreducible finite system 
size), and 
\beq
\llangle \C \vert \psi_R^{(0)} \rrangle 
= 
e^{- \beta E_\C} / Z 
\qquad 
Z \equiv \sum_\C e^{- \beta E_\C} 
. 
\eeq
[For convenience, in the following we shall assume that the relaxation rates 
are labelled in ascending order, 
$\varepsilon_0 = 0 < \varepsilon_1 \leq \varepsilon_2 \leq \ldots$] 
We can then decompose $\llvert P(t) \rrangle$ into normal modes, and use 
Eq.~\eqref{eq: Markov-eq-vec} to obtain 
$ \llvert P(t) \rrangle = \sum_n a_n(0) \, e^{- \varepsilon_n t} \, \llvert
\psi_R^{(n)} \rrangle $. In this language, we can express the ensemble
average of any given observable $\O$ as
\bea
\llangle \O(t) \rrangle_{\rm th} 
&=& 
\sum_\C \O_\C \, P_\C(t) 
\\ 
&=& 
\sum_\C \llangle \C \rrvert \hat{\O} \llvert P(t) \rrangle 
\equiv 
\llangle \Sigma \rrvert \hat{\O} \llvert P(t) \rrangle 
, 
\eea
where we defined $\llvert \Sigma \rrangle \equiv \sum_\C \llvert \C \rrangle$,
and $\hat{O} \llvert \C \rrangle = \O_\C \llvert \C \rrangle$.  Note that
the state $\llangle \Sigma \rrvert = \sum_\C \llangle\C\rrvert$ when bracketed 
with any probability state gives $\llangle \Sigma \vert P(t) \rrangle = 1$, 
i.e., the probability is normalised at all times, 
ensured by the fact that 
$\llangle \Sigma \rrvert =\llangle \psi_L^{(0)} \rrvert$ is a left eigenvector 
with zero eigenvalue. 
Decomposing into normal modes, we obtain 
\bea
\llangle \O(t) \rrangle_{\rm th} 
&=& 
\sum_n a_n(0) \, e^{- \varepsilon_n t} 
  \llangle \Sigma \rrvert \hat{\O} \llvert \psi_R^{(n)} \rrangle 
. 
\eea
If $\varepsilon_1$ remains finite in the thermodynamic limit, 
for $t \to \infty$ only the slowest mode $\varepsilon_0 = 0$ 
survives and one arrives at the thermodynamic equilibrium 
expression 
\bea
\llangle \O(t \to \infty) \rrangle_{\rm th} 
&=& 
\llangle \Sigma \rrvert \hat{\O} \llvert \psi_R^{(0)} \rrangle 
=
\sum_\C \O_\C \, \frac{e^{- E_\C / T}}{Z} 
. 
\nonumber 
\eea

We are interested in understanding in a precise way how the spectral 
properties of the operator $\hat W$ control the relaxation properties of the 
system. To this end, we investigate the two-time autocorrelation function: 
\bea
C(t+\tau,t) 
&=& 
\llangle \O(t+\tau) \, \O(t) \rrangle_{\rm th} 
\nonumber \\ 
&=& 
\sum_{\C,\Cp} 
  \O_{\Cp} \, P_{\C\to\Cp}(\tau) \: \O_\C \, P_\C(t) 
\nonumber \\ 
&=& 
\sum_{\C,\Cp} \llangle\C'| \hat\O e^{{\hat W} \tau} |\C\rrangle \;
\llangle \C| \hat\O |P(t)\rrangle
\nonumber \\ 
&=& 
\llangle\Sigma \rrvert \hat\O e^{{\hat W} \tau} \; \hat\O \llvert P(t) \rrangle
, 
\label{eq: autocorr}
\eea
where $P_{\C\to\Cp}(\tau)$ is the conditional probability that the system 
be in configuration $\Cp$ at time $\tau$, given that it was in configuration 
$\C$ at time $t=0$. 

Note that the right eigenvalues $\llvert\psi_R^{(n)}\rrangle$ of
$\hat{W}$ are not necessarily orthonormal. Thanks to the detailed
balance condition, it is possible to construct a real symmetric matrix
$\hat{H}_{\rm SMF}$ that has the same eigenvalues as $\hat{W}$ by means of a 
similarity transformation using the matrix 
$S_{\C,\Cp} \equiv \exp\left( -\beta E_\C / 2 \right) \, \delta_{\C,\Cp}$: 
\beq
\hat{H}_{\rm SMF} \equiv - \hat{S}^{-1} \hat{W} \hat{S} 
, 
\eeq
with 
$(H_{\rm SMF})_{\C,\Cp} = (H_{\rm SMF})_{\Cp,\C}
$ 
following from detailed balance. 

Labeling the eigenvectors of $H$ by $\llvert n \rrangle $, we take advantage of 
the fact that $\llangle n \vert m \rrangle = \delta_{n,m} $ and 
$\sum_n \llvert n \rrangle \llangle n \rrvert = \openone $ to express 
Eq.~\eqref{eq: autocorr} as 
\bea
C(t+\tau,t) 
&=& 
\llangle\Sigma\rrvert \hat{\O} \; e^{{\hat W} \tau} \; \hat{\O} \llvert P(t)\rrangle
\nonumber \\ 
&=& 
\llangle\Sigma\rrvert \hat{\O} \hat{S} \hat{S}^{-1} e^{{\hat W} \tau} 
  \hat{S} \hat{S}^{-1} \hat{\O} \llvert P(t)\rrangle
\nonumber \\ 
&=& 
\sum_n 
\llangle\Sigma\rrvert \hat{\O} \hat{S} e^{-{\hat H}_{\rm SMF} \, \tau} 
  \llvert n \rrangle \llangle n \rrvert \hat{S}^{-1} \hat{\O} \llvert P(t)\rrangle
\nonumber \\ 
&=& 
\sum_n e^{-\varepsilon_n \tau}
\llangle 0 \rrvert \hat{S}^{-1} \hat{\O} \hat{S} \llvert n \rrangle 
  \llangle n \rrvert \hat{S}^{-1} \hat{\O} \llvert P(t)\rrangle
, 
\nonumber \\ 
\eea
where we used the conservation of probability condition, namely that
$\llangle\Sigma\rrvert \hat{W} = 0$, to rewrite 
$\llangle\Sigma\rrvert = \llangle 0 \rrvert S^{-1}$. 
If we take the limit $t \to \infty$, 
\bea
C(\tau) 
&\equiv& 
\lim_{t\to\infty} C(t+\tau,t)
\nonumber \\ 
&=& 
\sum_n e^{-\varepsilon_n \tau}
\llangle 0 \rrvert \hat{S}^{-1} \hat{\O} \hat{S} \llvert n \rrangle 
  \llangle n \rrvert \hat{S}^{-1} \hat{\O} \llvert P(\infty)\rrangle
\nonumber \\ 
&=& 
\sum_n e^{-\varepsilon_n \tau}
\llangle 0 \rrvert \hat{S}^{-1} \hat{\O} \hat{S} \llvert n \rrangle 
  \llangle n \rrvert \hat{S}^{-1} \hat{\O} \hat{S} \llvert 0 \rrangle
, 
\eea
where we used the fact that 
$\llvert P(\infty) \rrangle=\hat{S} \llvert0\rrangle$.~\cite{footnote:q_EA} 
Classical thermodynamic observables are by definition measured in 
the preferred basis of classical configurations $\{\llvert\C\rrangle\}$. 
Therefore the operator $\hat\O$ is diagonal in this basis, it commutes with 
$\hat{S}$ ($\hat{S}^{-1}\; \hat\O\; \hat{S}=\hat\O$), and 
\beq
C(\tau) 
=
\sum_n e^{-\varepsilon_n \tau} \, 
\left\vert \llangle n \rrvert \hat{\O} \llvert 0 \rrangle \right\vert^2
. 
\label{eq: C(tau)}
\eeq
Notice that the r.h.s. in the above equation is precisely the GS 
quantum imaginary-time autocorrelation function $C^{\rm quant}(\tau)$ obtained 
from the operator $\hat\O$.~\cite{Henley2004} 

The connected part of the correlation function can then be written as 
\bea
C_c(\tau) 
&\equiv& 
\llangle \O(t+\tau) \, \O(t) \rrangle_{\rm th} 
- 
\llangle \O \rrangle^2_{\rm th} 
\\ 
&=& 
\sum_{n\ne 0} e^{-\varepsilon_n \tau} \, 
\left\vert 
  \llangle n \rrvert \hat{\O} \llvert 0 \rrangle 
\right\vert^2 
\\ 
&\equiv& 
C^{\rm quant}_c(\tau) 
. 
\nonumber 
\eea
%
%

\section{\label{sec: q_EA} 
Edwards-Anderson parameter 
        }
If we order the eigenvalues 
$\varepsilon_0 = 0 < \varepsilon_1 \leq \varepsilon_2 \leq \ldots$, we can 
write an upper bound of the connected autocorrelation function as 
\bea
C_c(\tau) 
&\le&
e^{-\varepsilon_1 \tau} \sum_{n \ne 0} \, 
\left\vert \llangle n \rrvert \hat{\O} \llvert 0 \rrangle \right\vert^2
\nonumber \\ 
&\le&
e^{-\varepsilon_1 \tau} \sum_{n} \, 
\llangle 0\rrvert \hat{\O} \llvert n \rrangle 
\llangle n \rrvert \hat{\O} \llvert 0 \rrangle
\\ 
&=& 
e^{-\varepsilon_1 \tau} \, \llangle 0\rrvert \hat{\O}^2 \llvert 0 \rrangle 
.
\label{eq: corr bound}
\eea

Unless $\varepsilon_1\to 0$ in the thermodynamic limit $N \to \infty$, 
the connected correlation function decays to zero exponentially fast in time 
($\tau_1=1/\varepsilon_1$), independently of the choice of the (bounded) 
observable $\O$. 
Therefore, any dynamical transition leading to `long' (i.e., non exponential) 
decay must be accompanied by a vanishing spectral gap in the SMF Hamiltonian. 

\begin{figure}[ht]
\begin{center}
\includegraphics[width=0.95\columnwidth]
                {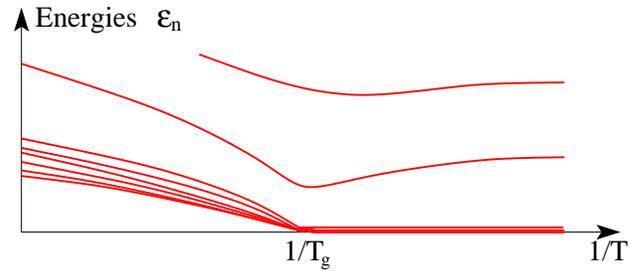}
\end{center}
\caption{
\label{fig: collapse} 
Qualitative illustration of the spectral collapse in the associated quantum 
system which is expected to occur at a glass transition in the original 
classical system with discrete degrees of freedom and no conserved quantities. 
}
\end{figure}

The appearance of vanishing relaxation rates in stochastic 
processes can be related to several different factors: 
for instance, critical slowing down at symmetry breaking phase transitions, 
Goldstone modes when a continuous symmetry is broken, or diffusive modes in 
paramagnetic phases with conserved quantities. 
Quite generally however these phenomena are considered distinct from 
glassiness in that they lead to parametrically large relaxation time scales 
(i.e., the gap closes as a power of system size). Glassiness on the contrary 
is characterised by a \textit{gap that vanishes exponentially in system size}. 
In this paper we focus on the latter phenomenon. 
Without loss of generality and for ease of discussion, 
we neglect slow modes that vanish as a power law of system size 
-- the reader can consider for example the case of a system with 
discrete degrees of freedom and no conserved quantities, 
illustrated qualitatively in Fig.~\ref{fig: collapse}. 
[Compare with a similar assumption in Ref.~\onlinecite{Gaveau1998}.] 
As a result, the high-temperature paramagnetic phase in the systems 
discussed below always relaxes exponentially fast in the thermodynamic 
limit, and the associated quantum Hamiltonian is in a gapped phase. 

Dynamical glassy phases are typically characterized by a 
finite value of the Edwards-Anderson order parameter, 
$q_{\rm EA} \equiv \lim_{\tau\to\infty} C_c(\tau)$.~\cite{Edwards1975} 
In quantum mechanical notation, 
\beq
q_{\rm EA}(\O) 
= 
\sum_{n\in {\D},n\ne 0} \, 
\left\vert 
  \llangle n \rrvert \hat\O \llvert 0 \rrangle 
\right\vert^2 
, 
\label{eq: EA parameter} 
\eeq
where $\D$ is the set of eigenvalues $\varepsilon_n$ that vanish in the limit 
of $N\to\infty$ (i.e., they collapse onto the GS). 

Let us postpone for the moment further considerations on how a non-vanishing 
$q_{\rm EA}$ arises in a quantum system, and let us look at its physical 
significance. 
As we pointed out earlier, the classical autocorrelation function in 
Eq.~\eqref{eq: C(tau)} is equivalent to the imaginary-time quantum 
autocorrelation function 
$
C^{\rm quant}(\tau) 
\equiv 
\llangle 0 \rrvert 
  e^{H_{\rm SMF}\,\tau}\hat{\O}e^{-H_{\rm SMF}\,\tau}\hat{\O} 
\llvert 0 \rrangle 
$. 
By temporally integrating the connected part of the quantum 
correlator, 
$
C^{\rm quant}_c(\tau) 
\equiv 
C^{\rm quant}(\tau) 
- 
\vert \llangle 0 \rrvert \hat{\O} \llvert 0 \rrangle \vert^2
$, 
one obtains the local static (zero-frequency) susceptibility of the quantum 
system at zero-temperature, 
\beq
\chi^{\rm loc} (\omega = 0) 
\equiv 
\int^\infty_0 \!\! d\tau \; C^{\rm quant}_c(\tau) 
=
\sum_{n \neq 0} 
\frac{\left\vert \llangle n \vert \hat{\O} \vert 0 \rrangle \right\vert^2}
     {\varepsilon_n} 
. 
\label{eq: Chi_loc}
\eeq

The Edwards-Anderson order parameter $q_{\rm EA}$, defined in 
Eq.~\eqref{eq: EA parameter}, corresponds to the long time limit 
$\tau\to\infty$ of the connected correlator $C^{\rm quant}_c(\tau)$. 
Therefore, the appearance of a non-vanishing value of $q_{\rm EA}$ 
coincides with the onset of a divergent local static 
susceptibility.~\cite{footnote:Griffiths} 

Notice that the susceptibility in Eq.~\eqref{eq: Chi_loc} measures the 
response of the quantum SMF Hamiltonian to a perturbation that couples to the 
operator $\hat{\O} = \sum_\C \vert\C\rangle\O_\C\langle\C\vert$. 
The value of $q_{\rm EA}$ is thus related to the second 
order correction in the GS energy of the perturbed system, 
$\hat{H}_{\rm SMF} + \lambda \hat{\O}$. 
We point out here that this perturbation is quantum mechanical in nature, 
and in general it is \emph{not} equivalent to some classical perturbation 
$E_\C\to E_\C+\delta E_\C$ in the original system. 

In conclusion, a dynamical transition in the original classical system 
becomes a well-defined thermodynamic quantum phase transition in the 
associated Hamiltonian $\hat{H}_{\rm SMF}$. 
This is a transition from a gapped to a gapless phase, detected by a 
non-vanishing order parameter ($q_{\rm EA}$), and accompanied by the 
appearance of a divergent static susceptibility. 
In other words, a 
{\it dynamical phase transition} can be precisely identified as an 
{\it equilibrium phase transition} 
in the {\it quantum system}. 

Conceptually the quantum perturbation that gives rise to the local static 
susceptibility discussed in this section is equivalent to those considered 
in large deviation functions.~\cite{large_dev_refs} 
Indeed, in the latter one perturbs the transition matrix $W$ that controls 
the dynamics of the classical process, $W \to W + sK$. The perturbation 
$K$ is intended to couple differently to slow and fast relaxing modes in the 
system. By tuning the coupling constant $s$ away from $s=0$, one thus 
favours/disfavours the low-lying eigenstates of $W$. 
Upon performing the similarity transformation in Sec.~\ref{sec: method}, 
$K$ maps onto a static perturbation $\tilde{K} = - S^{-1} K S$ to the quantum 
mechanical Hamiltonian $H = - S^{-1} W S$. 
So long as $K$ can be expressed as a local term in space and time in its 
classical formulation, the quantum operator $\tilde{K}$ is local, and one 
can \emph{qualitatively} understand the behaviour of the system as a function 
of the coupling constant $s$,~\cite{large_dev_firstorder_refs} as follows. 
At high temperature the quantum system is gapped (see above for a discussion 
of this assumption). Therefore, varying $s$ can induce a transition in the 
system only at finite values of $s$ (if at all present). 
On the other end, below a glass transition, there is a large number of 
degenerate lowest-lying states. At least in finite size systems, the 
degeneracy is not exact, and the operator $\tilde{K}$ is likely to split 
the degeneracy, inducing a first order transition at $s=0$. 
In order to make this argument rigorous, the thermodynamic limit ought to be 
properly accounted for, which is beyond the scope of the present paper. 
(Note that this behaviour occurs also at classical symmetry breaking 
transitions.) 
%
%
%

\section{\label{sec: fidelity} 
Fidelity susceptibility
        }
The information theoretic concept of fidelity has been recently 
used to detect quantum phase transitions without \textit{a priori} knowledge 
of any order paramater.~\cite{fidelity_refs,fidelity_susc_refs} 
It has also proven useful to investigate the nature of a critical point, 
allowing one to derive scaling exponents and other universal properties. 
This technique provides a general approach to study a dynamical phase 
transition, that encompasses the case where no local order parameter is found 
even at the quantum level 
(e.g., topologically ordered systems~\cite{Abasto2008}). 

The fidelity overlap is defined as 
\bea
\F(\beta,\delta\beta) 
&\equiv& 
\llangle \psi_0(\beta-\delta\beta/2) 
  \,\vert\, 
\psi_0(\beta+\delta\beta/2) \rrangle 
, 
\eea
from which one obtains the so-called fidelity susceptibility 
\bea
\chi_\F(\beta) 
&\equiv& 
\lim_{\delta\beta \to 0} 
\left[ 
  -2 \frac{\ln \F(\beta,\delta\beta)}{\delta\beta^2} 
\right] 
. 
\eea

One of the features of the classical to quantum correspondence is that 
the GS wavefunction of the quantum system is known exactly for all values 
of $\beta$, which allows for a direct calculation of $\chi_\F(\beta)$. 
For a generic SMF Hamiltonian, the GS wavefunction can be written as 
\beq
\llvert \psi_0(\beta) \rrangle 
= 
\sum_\C \frac{\exp(-\beta E_\C / 2)}{\sqrt{Z(\beta)}} \, \llvert\C\rrangle 
, 
\label{eq: SMF-wf}
\eeq
where $ Z(\beta)=\sum_{\C} \exp(-\beta E_{\C}/2)$ is the partition
function of the classical model. 
{}From it, we obtain that the fidelity susceptibility $\chi_\F$ of the SMF 
quantum system is proportional to the heat capacity $C_V$ of the 
original classical system, 
\bea
\chi_\F(\beta) 
&=& 
\frac{1}{4} \frac{d^2}{d\beta^2}\ln Z(\beta) 
\nonumber \\ 
&=& 
\frac{1}{4\beta^2} C_V(\beta) 
, 
\eea
where we used the fact that $E = -d\ln Z(\beta)/d\beta$, and that 
$C_V = dE / dT$ is the heat capacity of the classical system.

Singularities in the heat capacity (not necessarily a divergence) are 
therefore in one-to-one correspondence with singularities 
in the GS fidelity susceptibility.  
Whilst there is no rigorous proof that all quantum phase transitions in 
systems with local Hamiltonians give rise to singularities in $\chi_\F$, 
no counter examples are known, including quantum disordered and topologically
ordered systems.~\cite{footnote:p2spin_model} 
This is suggestive that a generic dynamical phase transition in a \emph{local} 
classical system is detected by the fidelity susceptibility as a thermodynamic 
transition in the associated quantum system, and it is likewise signalled by 
a \emph{necessary} singularity in the heat capacity of the original classical 
system. 
%
%

\section{\label{sec: entanglement}
Entanglement entropy to probe glass transitions and diverging length scales 
        }
The von Neumann entanglement entropy has recently been applied quite 
extensively as a tool to study quantum systems.~\cite{Amico-etal} 
It provides an unbiased measure, in the sense that it does not
focus on a particular order parameter or local operator, to probe
properties that are not accessible via standard correlation functions
among a fixed number of degrees of freedom. Because it is not hinged on order 
parameters, the entanglement entropy can be a means to uncover 
``hidden'' order that is not easily detectable otherwise. For example, 
it has been used to detect topological orders in systems 
where no local order parameters 
exist.~\cite{Hamma2005,Levin2006,Kitaev2006} 

One measure of entanglement is block entanglement, obtained by
bipartitioning the system into two blocks $A$ and $B$, and computing
the von Neumann entropy $S_{AB}$ for the reduced density matrix after
tracing out the degrees of freedom in $B$ (or $A$). When the system is
short range correlated, the entanglement entropy $S_{AB}$ obeys the
so-called area law,~\cite{Srednicki} {\it i.e.}, 
$S_{AB}\propto \ell_{AB}^{d-1}$, where $\ell_{AB}$ is the length scale of the
boundary between the subsystems and $d$ is the dimension of space. At
criticality, it is possible to have logarithmic corrections to the
area law, as is the case in one dimension (1D).~\cite{Calabrese-Cardy2004} 
However, this is not always the case in higher dimensions, and it is possible
to have no corrections to the area law even when the correlation
length diverges.~\cite{Verstrate-etal2006} 
A correction to the area law near a critical point is not the only signature 
of a quantum phase transition. In the case of a topological phase
transition, subleading corrections to the entanglement entropy that
capture the topology of the bipartitions do show a clear change as
the system crosses transitions between topological and
non-topological phases.~\cite{Castelnovo2008a} 
Moreover, one can use the scaling of the (topological) entanglement entropy 
as a function of the size of the bipartitions to define a growing 
correlation length that diverges at the trasition, even if the system is 
devoid of a local order parameter. 

Here we propose that the entanglement entropy for the mapped quantum
system can be used as a tool to study glass transitions that are hard
to probe otherwise, say, using equal-time correlation 
functions. Although the entanglement entropy is generally difficult to
compute, the point that we make in this section is that the behavior
of the entanglement entropy can provide a formal way to uncover glass
transitions without \textit{a priori} knowledge of any order parameter, and 
without the need of ad hoc definitions of time-scale-dependent free energy 
basins. 
We remark that computing the entanglement entropy amounts to a static
calculation in the quantum mechanical language. Nonetheless, as we
discuss below, it captures the properties of metastable states
in the original system, which can only be defined by a 
distinction between ``short'' and ``long'' time scales in the
classical time-dependent language.

\begin{figure}[ht]
\begin{center}
\includegraphics[width=0.5\columnwidth]
                {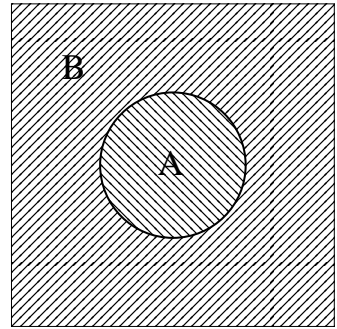}
\end{center}
\caption{
\label{fig: AandB} 
Bipartition of a system into subsystems $A$ and $B$. 
}
\end{figure}

For concreteness, consider a standard bipartition of the degrees of
freedom of the system into two subsets: subsystem $A$, a bubble of finite 
radius $R$, and subsystem $B$, as shown in Fig.~\ref{fig: AandB}. 
Using the density matrix constructed from the GS 
wavefunction Eq.~\eqref{eq: SMF-wf}, 
\begin{eqnarray}
\label{eq:rho}
\hat\rho(\beta)
&=&
|\psi_0(\beta) \rangle 
\langle \psi_0(\beta)|
\nonumber\\
&=&
\frac{1}{Z(\beta)}
\sum_{\C,\C'} 
e^{-\beta (E_\C+E_{\C'})/2}
\, |\C\rangle \langle \C'|
\;
\end{eqnarray}
and tracing over subsystem $B$ to get the reduced density matrix
$\hat\rho_A(\beta)={\rm Tr}_B \hat\rho(\beta)$, one obtains
the entanglement entropy 
${S}_{AB}(\beta)=-{\rm Tr}_A[\hat\rho_A(\beta)\ln \hat\rho_A(\beta)]$. 
The entaglement entropy for a pure state, like the one in 
Eq.~\eqref{eq: SMF-wf}, is  symmetric, 
i.e., one arrives at the same result by computing 
$S_{AB} = -{\rm Tr}_B[\hat\rho_B(\beta)\ln \hat\rho_B(\beta)]$. 

The computation of $S_{AB}$ from Eq.~\eqref{eq:rho} was carried out explicitly 
in Sec.~V of Ref.~\onlinecite{Castelnovo2007a}: 
\bea
S_{AB} (\beta)
&=& 
\beta F^{\ }_{A} 
+ 
\beta F^{\ }_{B} 
- 
\beta F^{\ }_{A \cup B} 
+ 
\beta \llangle E^{\partial}_{\ } \rrangle_{\rm th} 
\eea
where 
$\beta F^{\ }_{A} = - \ln Z^{A}_{D}$, 
$\beta F^{\ }_{B} = - \ln Z^{B}_{D}$, and 
$\beta F^{\ }_{A \cup B} = - \ln Z$. 
The partition functions $Z^{A}_{D}$ and $Z^{B}_{D}$ are computed with
Dirichlet boundary conditions on the interface between the two
subsystems, {\it i.e.} they are computed by pinning the configurations
at the boundary of the bipartition and running the sums over
configurations in the bulk of $A$ or $B$, respectively. That is to be
contrasted with $Z^{A}_{F}$ and $Z^{B}_{F}$, where the partition
function is computed with free boundary conditions, simply summing
over all configurations in $A$ or $B$ without boundary constraints. 
Notice that the total partition $Z(\beta) \equiv Z^{A\cup B}_{F}$ is 
naturally computed with free boundary conditions, as there is no boundary 
in the total system $A\cup B$. 
Finally, $\llangle E^\partial\rrangle_{\rm th}$ is the 
average energy of the interface between $A$ and
$B$ (i.e., the sum of all terms in $E_{\C}$ that involve simultaneously 
degrees of freedom in $A$ and in $B$) 
computed with the measure $ P(\C)=\exp(-\beta E_{\C})/Z(\beta)$.

It is instructive to rewrite the boundary entropy in the following
form,
\bea
S_{AB} (\beta)
&=& 
\beta (F^{\ }_{A} -F^{F}_{A})+\beta (F^{\ }_{B} -F^{F}_{B})
\nonumber\\
&+& 
\beta F^{F}_{A} +\beta F^{F}_{B} 
- 
\beta F^{\ }_{A \cup B} 
+ 
\beta \llangle E^{\partial}_{\ } \rrangle_{\rm th} 
. 
\eea
We can then define the free-boundary-condition part of the entanglement 
entropy 
\begin{eqnarray}
S^F_{AB} (\beta)
&\equiv&
\beta 
  \left(\vphantom{\sum}
    F^\partial
    -
    \llangle E^{\partial}_{\ } \rrangle_{\rm th} 
  \right)
\end{eqnarray}
where 
$
F^\partial
\equiv 
F^{\ }_{A \cup B} 
- F^{F}_{A} 
- F^{F}_{B} 
= 
\beta^{-1}
\ln\llangle \exp{\left(\beta E^{\partial}_{\ } \right)}\rrangle_{\rm th}
$, 
and we can introduce the notation 
\begin{eqnarray}
\Delta F_{A,B} 
\equiv 
F^{\ }_{A,B} -F^{F}_{A,B} 
= 
-\frac{1}{\beta} 
\ln \frac{Z^{A,B}_{D}}{Z^{A,B}_{F}}
\;,
\end{eqnarray}
so that 
\begin{eqnarray}
S_{AB} (\beta)=\beta\left(\Delta F_{A}+\Delta F_{B}\right)+S^F_{AB} (\beta) 
. 
\label{eq: S_AB rewritten}
\end{eqnarray}

The term $S^F_{AB}(\beta)$ can identified with the cumulant-generating
function for the fluctuations of the boundary energy,
\begin{eqnarray}
S^F_{AB} (\beta) 
= 
\ln 
\llangle
  e^{\beta\left(
    E^{\partial}_{\ }-\langle E^{\partial}_{\ }\rangle_{\rm th} 
  \right)} 
\rrangle_{\rm th} 
.
\label{eq: S_AB cumulant}
\end{eqnarray}
The second cumulant is a measure of the heat capacity for the boundary 
degrees of freedom. $S^F_{AB}(\beta)$ generically obeys the area 
law away from quantum critical points. If there is a singularity in 
the heat capacity of the system, then this term becomes singular 
and picks up the phase transition, much like the fidelity susceptibility in 
Sec~\ref{sec: fidelity}. 

However, as we argue in the following, the entanglement entropy $S_{AB}$ 
in Eq.~\eqref{eq: S_AB rewritten} is able to detect glass transitions 
even in subtle cases when there are truly no singularities in the specific 
heat / quantum fidelity. 
This can be seen by considering the other terms in $S_{AB}$, namely 
$\Delta F_{A,B}$, which contain information on the effects of 
the boundary conditions on the free energy of subsystems $A$ and $B$. 


Let us adopt a description of the phase space of 
subsystem $A$ in terms of metastable free energy basins, commonly used in the 
literature of glassy phenomena. 
The goal here is to 
argue that the von Neumann entropy exhibits signatures of glassy phenomena, 
and it can be used to determine growing length scales and other 
characteristic features. 
On the other hand, however, such time-scale-dependent description in terms 
of metastable states is not needed to compute / study the von Neumann entropy 
in the quantum mechanical language, and we argue that $S_{AB}$ 
provides a static measure to investigate those same glassy phenomena, 
without the need to justify a metastable state description of the system. 

Consider the bipartition $(A,B)$ of the system illustrated in 
Fig.~\ref{fig: AandB}. 
Let us first allow the whole system to equilibrate, and then let us freeze 
subsystem $B$ and let $A$ 
thermalize with fixed boundary conditions. 
In the high-temperature liquid phase the fixed boundary conditions play a 
marginal role, and the free energy of $A$ is characterised by a single 
minimum, which is not significantly different from the one that obtains for 
free boundary conditions. 
On the other hand, glassiness is typically accompanied by the appearance 
of distinct free energy minima, and the partition function of $A$ can thus 
be decomposed accordingly into separate contributions. 
Following Bouchaud and Biroli~\cite{Bouchaud2004}, we make the coarse 
assumption that all these contributions have the same bulk free energy, 
and differ only by a boundary energy term. 
Moreover, we take this boundary term to be the same for all 
contributions ($E_\partial^*$) but for one that is favoured 
($E_\partial^*-\Delta E$) by an amount $\Delta E > 0$. 
If we label these minima in the free energy of $A$ by indices $\alpha_i$, 
$i=1,\dots,{\cal N}$, with  interfacial energy 
$E_{\alpha_{i\ne 1}}\sim E_\partial^*$, and 
$E_{\alpha_{1}}\sim E_\partial^*-\Delta E$, then 
under the phenomenological assumptions introduced above it is possible to
write
\begin{eqnarray}
\Delta F_{A}
&=&
-\frac{1}{\beta}
\ln\frac{Z^{A}_{D}}{Z^{A}_{F}}
\nonumber\\
&\approx&
-\frac{1}{\beta}
\ln
\left[
\frac{
\sum_{i=1}^{\cal N} e^{-\beta E_{\alpha_{i}}}
}
{\cal N}
\right]
\nonumber\\
&\approx&
E^*_\partial
-\frac{1}{\beta}
\ln
\left[
\frac{
({\cal N}-1)+
\; e^{\beta \Delta E}
}
{\cal N}
\right]\nonumber\\
&\approx&
E^*_\partial
-\frac{1}{\beta}
\ln
\left[
1+e^{\beta \Delta E-S^*}
\right] 
, 
\label{eq:comp-BB}
\end{eqnarray}
where $S^*=\ln {\cal N}$ is the configurational entropy of $A$. 

{}From Eq.~\eqref{eq:comp-BB}, we can obtain the behavior of the
contribution $\Delta F_{A}$ to the von Neumann entanglement entropy of
the associated quantum system across the glass transition
temperature. Let us focus on the case where the configurational
entropy $S^*$ becomes extensive in the glassy phase. 
Above the glass transition, the liquid phase (where ${\cal N} = 1$, and 
$S^* = 0$) quickly relaxes to minimize 
the energy strain due to the fixed boundary conditions, 
and $\Delta F_{A} \sim E_{\alpha_1} = E^*_\partial - \Delta E$. 
Below the transition, the appearance of long relaxation time scales 
prevents the system from moving between the ${\cal N}$ minima. 
For large enough $R$, the extensive entropy $S^* \sim R^d$ 
dominates the exponential in Eq.~\eqref{eq:comp-BB}, and 
$\Delta F_{A} = E^*_\partial$. 

\begin{figure}[ht]
\begin{center}
\includegraphics[width=0.8\columnwidth]
                {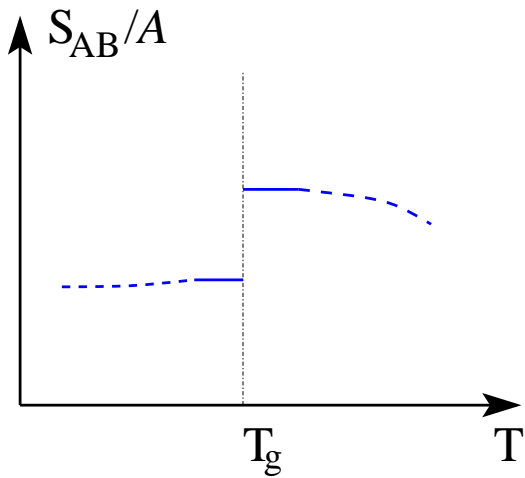}
\end{center}
\caption{
\label{fig: S_vN behaviour 1} 
Qualitative illustration of the behaviour of $S_{AB} / \mathcal{A}$ 
in the thermodynamic limit as a function of $T$, if we take, say, 
subsystem $A$ to be half of the entire system. 
}
\end{figure}

This behaviour becomes evident if we take 
$A$ to be half of the entire system, and we consider the thermodynamic limit. 
In this case, $\Delta F_A = \Delta F_B$. 
Irrespective of $S^*$ being finite or extensive, 
the area law is obeyed ($\Delta F_A\propto R^{d-1}$), but with 
{\it different} coefficients (namely, $(E^*_\partial - \Delta E)/R^{d-1}$ 
vs $E^*_\partial/R^{d-1}$). 
Therefore, at least within the scenario 
where the number of minima scales extensively with (sub)system size, 
the glass transition can 
be captured by a sudden change in slope of the von Neumann entropy of a 
bipartition of the system as a function of the interface area, 
as illustrated in Fig.~\ref{fig: S_vN behaviour 1}. 
(Note that this conclusion is reached within a mean field picture that 
neglects possible additional logarithmic corrections to the area law 
at the transition temperature.) 

We stress that, although we used a metastable state description 
to obtain the behaviour of $\Delta F_A$, the entanglement entropy is a 
{\it static} measure, that is defined in terms of the quantum ground state 
of the mapped Hamiltonian, i.e, it is related to the classical 
{\it equilibrium} state. 
Therefore, the entanglement entropy can be used independently 
of whether we have a proper understanding of the metastable states in the 
classical system. 

\begin{figure}[ht]
\begin{center}
\includegraphics[width=0.8\columnwidth]
                {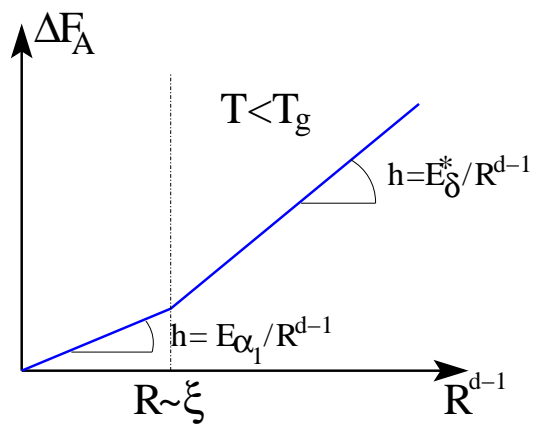}
\end{center}
\caption{
\label{fig: S_vN behaviour 2} 
Qualitative illustration of the behaviour of $\Delta F_{A}$ as a function of 
area $\mathcal{A}(R)$, as predicted in the scenario discussed in 
Ref.~\onlinecite{Bouchaud2004} below the glass temperature $T_g$. 
}
\end{figure}

In addition to its behavior across the transition, the dependence 
of the von Neumann entanglement entropy on the size of the 
bipartition can be used to detect a growing correlation length $\xi$ 
that diverges at the transition. 
Here, as in Ref.~\onlinecite{Bouchaud2004}, $\xi$ is identified as the 
crossover size of $A$ such that $S^*\sim\beta\Delta E$ 
(recall that $S^*\sim R^d$ and $\Delta E\sim R^{d-1}$). 
The value of $\xi$ depends on the temperature both explicitly ($\beta$) 
and through the (intensive) configurational entropy $s^* = S^*/R^d$. 
Within the metastable state picture, 
it follows from Eqs.~\eqref{eq: S_AB rewritten}
and~\eqref{eq:comp-BB} that for $R\ll\xi$ one finds $\Delta F_{A} \sim
E^*_\partial - \Delta E = E_{\alpha_1}$ (i.e., the energy difference
dominates over the entropic gain), 
while for $R \gg \xi$ one finds $\Delta F_{A}\sim E^*_\partial$ (i.e.,
the entropy dominates, and the free energy difference between the
fixed and free boundary conditions is given by the boundary energy of
the majority of the configurations of $A$). Thus, a plot of $\Delta
F_{A}$ (and therefore of $S_{AB}$) as a function of the interface area
$\mathcal{A}(R) \sim R^{d-1}$ exhibits a kink around $R \sim \xi$, where the 
slope changes from $E_{\alpha_1}/\mathcal{A}$ to $E^*_\partial/\mathcal{A}$ 
(see Fig.~\ref{fig: S_vN behaviour 2}). 
%
%


We would like to stress the following important point. The length
scale defined by Bouchaud and Biroli required a careful definition of
the metastable states, based on the existence of a regime of time
scales sufficiently large compared with the equilibration time within
each metastable state, but small compared to transitions between the
states. Again, here we used their results just to argue that the
entanglement entropy does capture this growing length scale, and
exhibits a distinctive behaviour across the transition. However, the
entanglement entropy per se {\it does not} require {\it any}
discussion of metastable states and time scales. The entanglement
entropy computed from Eq.~\eqref{eq: S_AB rewritten}
is a {\it static} measure of the 
{\it equilibrium} state of the system. Our point is that this static but
non-local measure can be used to probe/reveal a ``hidden'' static
transition, and the associated growing correlation length.

In summary, we propose that the entanglement entropy can be used as a
way to probe glass phases and phase transitions without bias towards
order parameters. Within the scenario used for instance in
Ref.~\onlinecite{Bouchaud2004}, we showed that the entanglement
entropy obeys the area law, with a prefactor that changes abruptly across 
the glass transition. It can also be used to identify static growing
length scales that in the classical language require a time-dependent
formulation in terms of configurational
entropies,~\cite{Biroli2001,Bouchaud2004} or point-to-set correlation
functions.~\cite{Montanari2006} The entanglement entropy that we 
propose to use is a completely static measure that can serve as a 
probe of non-local order, and thus it can be a tool to reveal an 
underlying static transition associated with the dynamical glass 
transition. 
%
%

\section{\label{sec: KCM example}
Hidden order in an ideal glass plaquette model
        }
In this section we discuss a class of models 
with short-range interactions 
that exhibit a large number of degenerate lowest energy 
configurations, and we consider the nature of the energy barriers between 
such configurations at low temperatures. 
In particular, for one of the models we show that the existence of a 
thermodynamic phase transition can be argued by mapping it to essentially 
decoupled 2D Ising layers. 
Below the phase transition, the energy barriers 
between the many energy minima grow exponentially with system size, 
and the existence of an equal number of slow relaxing 
modes can be shown rigorously by means of a variational calculation involving 
the associated SMF Hamiltonian. 
%
%

\subsection{\label{sec: generalised gonihedric model}
Generalised gonihedric model
           }
Consider Ising spin variables $\s_\ii$ defined on
the sites $\ii\equiv(i_x,i_y,i_z)$ of a cubic $L\times L\times L$ lattice
(with periodic boundary conditions imposed), 
with energy~\cite{Koutsoumbas2002} 
\begin{eqnarray}
E &=& -J_{xy}\sum_\ii 
\s_{\ii}\,\s_{\ii+\hat\xx}\,\s_{\ii+\hat\xx+\hat\yy}\,\s_{\ii+\hat\yy}
\nonumber \\ 
&& -J_{yz}\sum_\ii 
\s_{\ii}\,\s_{\ii+\hat\yy}\,\s_{\ii+\hat\yy+\hat\zz}\,\s_{\ii+\hat\zz}
\nonumber \\ 
&& -J_{zx}\sum_\ii 
\s_{\ii}\,\s_{\ii+\hat\zz}\,\s_{\ii+\hat\zz+\hat\xx}\,\s_{\ii+\hat\xx}
\;,
\label{eq: generalised gonihedric sigma}
\end{eqnarray}
where $\hat\xx$, $\hat\yy$ and $\hat\zz$ are the unit vectors of the cubic 
lattice. 
The system contains interactions among four spins at the vertices of
square plaquettes, with coupling constants that depend on whether a
plaquette is parallel to the $xy$, $yz$, or $zx$ plane. We 
focus on the case where 
\bea
&& 
J_{yz}=J_{zx}=J 
\\ 
&& 
J_{xy}=J^\prime 
, 
\eea
which has an anisotropy between the coupling constants for vertical 
and horizontal plaquettes (with respect to the $xy$ plane). 

Two particular limiting cases of this model are equivalent to systems
already studied in the literature. 
For $J=0$, $J^\prime \neq 0$ the system is equivalent to $L$ 
decoupled copies of the 2D square plaquette 
model.~\cite{2Dplaquette_refs,Savvidy1996} 
This model is known to exhibit an activated behaviour, with time scales 
growing as the exponential of the inverse temperature.~\cite{Jack2005} 
The model does not exhibit a dynamical transition, and the longest relaxation 
time scale diverges only in the zero temperature limit. 

The isotropic limit $J=J^\prime$ corresponds to the gonihedric 
model.~\cite{Johnston2007} 
This model has been argued to undergo a first order 
thermodynamic transition, using numerical simulations and cluster mean field 
arguments.~\cite{gonihedric_first_order} 
Moreover, numerical simulations point to the existence 
of a finite temperature glass transition in the neighbourhood of the 
thermodynamic transition.~\cite{gonihedric_glass} 
However, no analytical approach has been successful at confirming the 
presence of such dynamical transition thus far. 

The case we shall consider here is the limit $J\ne 0$, $J^\prime=0$, which 
we refer to as the anisotropic gonihedric 
model. 
As we show below, this model has a thermodynamic transition at 
$T_c = 2 J/\log(1+\sqrt{2})$, 
which is the value of the 
2D Ising transition temperature for a nearest-neighbor interaction 
$J$  (see Refs.~\onlinecite{anisotropic_gonihedric_refs} 
and~\onlinecite{Koutsoumbas2002} for a study of the system based on 
transfer matrix considerations). 
In contrast with a single Ising plane that has two degenerate 
minima, this system has a number of degenerate energy minima that scales 
as $2^{L-1}$. 
%
%

\subsubsection{\label{sec: J'=0 gonihedric model}
$J^\prime=0$ -- kinetically constrained model with a finite 
temperature transition
              }
\begin{figure}[ht]
\begin{center}
\includegraphics[width=0.6\columnwidth]
                {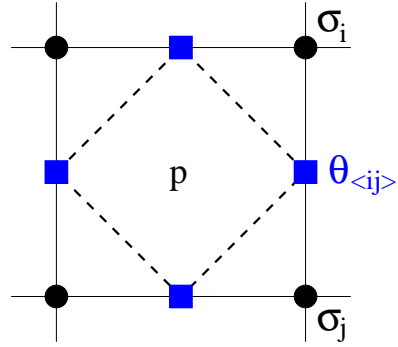}
\end{center}
\caption{
\label{fig: sigma2theta plaquette} 
Illustration of the mapping from $\sigma$ (solid circles) to $\theta$ 
(solid squares) spins, for a plaquette $p$ of the cubic lattice. 
}
\end{figure}

Let us write the anisotropic gonihedric model as a gauge
theory, introducing Ising degrees of freedom $\theta_{\langle \ii \jj
\rangle}$ defined on the links (or bonds) ${\langle \ii \jj
\rangle}$ between nearest neighbor sites $\ii$ and $\jj$ of the cubic
lattice (see Figure~\ref{fig: sigma2theta plaquette}),
\begin{equation}
\theta_{\langle \ii \jj \rangle} = \s_{\ii}\,\s_{\jj} 
. 
\end{equation}
The new Ising degrees of freedom are subject to the hard constraint that 
the product of four $\theta$ spins along the edges of any given plaquette must 
be equal to 1 (gauge constraint). 
Under periodic boundary conditions, there are additional constraints on the 
$\theta$ spins, since products along lines winding around the system (say, 
parallel to the $x$, $y$, or $z$ axis) must also equal $1$. 

Because of the gauge constraint, one can actually work solely with
$\theta$ variables on the vertical bonds between sites $\ii$ and
$\ii+\hat \zz$, with centers at $\bb \equiv (i_x,i_y,i_z+1/2)$:
\begin{equation}
\theta_{\langle \ii \jj \rangle}
\equiv
\theta_{\bb}
 = \s_{\bb-\hat\zz/2}\,\s_{\bb+\hat\zz/2} 
. 
\label{eq:vertical}
\end{equation}
There is a $1$-to-$1$ mapping between the original $\sigma$ spin 
configurations, and the (constrained) configurations of $\theta_\bb$ spins,
provided that we specify the values of the $\sigma_\ii$ spins on a given 
reference $xy$ plane, e.g., $\ii \equiv (i_x,i_y,0)$. 
In the limit of interest, $J \neq 0$ and $J^\prime = 0$, the energy of the 
system can be expressed purely in terms of vertical $\theta_\bb$ spins, 
\bea
E = -J \sum_{\bb} 
\left[
  \theta_{\bb}\:\theta_{\bb+\hat{\xx}}
  +
  \theta_{\bb}\:\theta_{\bb+\hat{\yy}}
\right] 
. 
\label{eq: anisotropic gonihedric theta energy}
\eea
The plane of $\sigma$ spins required by the mapping behaves as a 
separate paramagnetic contribution to the system, which can be 
disregarded. 

In the $\theta$ spin language, the limit $J^\prime=0$ of the generalised 
gonihedric model in Eq.~\eqref{eq: generalised gonihedric sigma} can be 
recognised as a collection of decoupled 2D Ising models in disguise 
[see Eq.~\eqref{eq: anisotropic gonihedric theta energy}]. 
This is not a trivial result, and in the original $\sigma$ spin 
formulation the $J^\prime=0$ model, while being clearly anisotropic, is far 
from being factorisable into decoupled 2D layers. 

Note that the decoupling is not exact. Indeed, the product of theta spins 
along vertical lines $\prod^L_{i_z=1} \theta_{\bb = (i_x,i_y,i_z+1/2)}$ must 
equal one because of the nature of the mapping (gauge condition). 
However, we do not expect the parity constraint to affect the properties of 
the system in the thermodynamic limit, as it simply reduces the number of 
independent $\theta$ degrees of freedom by $L^2$. 

Below the 2D Ising transition temperature 
$T_c=2J/\log(1+\sqrt{2})$ the system orders, and each layer 
(which we can label by $k=1,\dots,L$) acquires an (independent) expectation 
value $M_k = \langle\theta_{(i_x,i_y,k+1/2)}\rangle = \pm M(T/T_c)$ 
(apart from the parity constraint, enforcing 
$\prod^L_{i_z=1} \theta_{\bb = (i_x,i_y,i_z+1/2)} = +1$). 
There are therefore $2^{L-1}$ minima that correspond to all relative 
magnetizations of the layers. 
This is an example of a system without disorder but with a true thermodynamic 
transition into a phase with many minima, whose number scales subextensively 
with system size (namely, exponentially in $L = \sqrt[3]{N}$). 

In a 2D Ising ferromagnet below $T_c$, the system settles in one of the two 
minima (spontaneous symmetry breaking), and the energy barrier between them 
grows with the linear size of the system. 
As a result, the time scale for the system to migrate from one minimum to the 
other grows exponentially in $L$ (see for instance 
Appendix~\ref{app: variational excited states}). 
The same is true for the system in question, except that the number of minima 
separated by energy barriers $\sim\!L$ scales as $2^L$. We therefore expect to 
observe an equal number of independent slow relaxation modes in the 
transition matrix of the system (assuming local $\sigma$ spin dynamics), 
with rates $\sim \exp(-L)$, which vanish in the thermodynamic limit. 
In Sec.~\ref{sec: KCM example - variational states} we explicitly show the 
existence of these slow modes by means of a variational approach in the 
associated SMF quantum language. 

Whereas the $\theta$ language allows one to reduce the system to decoupled 2D 
Ising planes, simple dynamical processes in the $\sigma$ spins translate into 
coordinated rearrangements in the $\theta$ spins -- a feature that is 
characteristic of kinetically constrained models. 
In the present case, local dynamics in the $\sigma$ spins are bound to couple 
the dynamics of different $\theta$ planes, reflecting the fact that the system 
is in truth three dimensional. 
For instance, single $\sigma$ spin flip events translate into flipping two 
vertically adjacent $\theta$ spins, that is, spins that belong to two 
adjacent planes: 
\bea
\s_{\ii}
\to 
-\s_\ii
\quad\Rightarrow\quad 
\begin{cases}
  \theta_{\ii-\hat{\zz}/2} \to -\theta_{\ii-\hat{\zz}/2}
  \\
  \theta_{\ii+\hat{\zz}/2} \to -\theta_{\ii+\hat{\zz}/2}
\end{cases} 
. 
\label{eq: dynamics-theta}
\eea
This is an example of a kinetically constrained model with a 
{\it finite} temperature thermodynamic transition. 
The coordinated dynamics can affect the coarsening taking place within each 
2D layer of $\theta$ spins. 
For example, in order for a domain wall in one of the layers to move, 
$\theta$ spins must be flipped either in the layer immediately above or below, 
as a consequence of Eq.~\eqref{eq: dynamics-theta}. 
If these spins happen to be within an ordered domain, as opposed to a 
boundary between domains, they induce an energy cost to the domain wall motion 
in the adjacent layer, and therefore coarsening becomes an activated process. 

We can summarize the features of this model as follows:
\\ 
(I) A $\s$ spin can be written as a product of vertical $\theta$ spins
and a $\s$ in the reference plane. The average of the product scales
as $[\pm M(T/T_c)]^d$, where $M(T/T_c) \in (0,1)$ is the magnitude of the 
average magnetisation of a spin, and $d$ is the distance to the reference 
plane. Therefore, $\langle \sigma \rangle\to 0$ exponentially fast 
away from the reference plane. The system is a spin 
liquid, with no long-range order in the $\s$-spins. 
[See also Ref.~\onlinecite{Koutsoumbas2002}, where it is shown using transfer 
matrix arguments that the correlation length of the system remains finite 
across the transition.] 
\\ 
(II) Nevertheless, 
$
\langle \s_\ii \s_{\ii+\hat{\zz}} \rangle 
=
\langle\theta_{\ii+\hat\zz/2}\rangle 
= 
\pm M(T/T_c)
$, 
$\forall\,i_z=1,\ldots,L$, 
and there are sub-extensively many minima ($2^{L-1}$). 
Provided that the dynamics in the $\sigma$ spins are local, the system is 
characterised by an equal number of slow relaxation modes, whose rates 
vanish exponentially $\sim \exp(-L)$ in the thermodynamic limit. 
\\ 
(III) In addition to these exponentially large time scales, 
the two-time average 
$
\langle \s_\ii(t) \,\s_{\ii+\hat\zz}(t) 
  \,\s_\ii(t_w) \,\s_{\ii+\hat\zz}(t_w) \rangle 
$ 
($= \langle\theta_{\ii+\hat\zz/2}(t)\,\theta_{\ii+\hat\zz/2}(t_w)\rangle$) 
shows at least $t/t_w$ scaling (coarsening within the planes), if not an even 
slower 
aging behaviour due to the coordinated inter-plane dynamics. 

%
%

\subsubsection{
$J^\prime\ne 0$ -- the relation to the kinetically constrained 
square plaquette model, and confinement
              }
Suppose that one starts from the limit $J^\prime \ne 0$, and $J=0$, 
i.e., truly decoupled 2D square plaquette layers. As was shown in 
Ref.~\onlinecite{Jack2005}, it is well understood that the dynamics 
become slow as temperature is reduced. 
Single $\sigma$ spin flip updates change the sign of all four adjacent 
plaquettes in the plane, so in order to move an isolated defective 
plaquette (one where the product of the four $\s$ spins at the corners is 
negative) requires the creation of extra pairs of defective plaquettes, 
each at a cost $J^\prime$. 
However, pairs of defective plaquettes can move freely across the system, 
and defects can therefore be annihilated or separated by processes that 
encompass only finite energy barriers, repeated for a number of steps that 
scales linearly with the distance between the defects. 
As a result, the system exhibits a conventional activated behaviour, and the 
time scales diverge only in the zero temperature limit. 
In other words, defects in the 2D square plaquette model are 
\emph{deconfined}, as only a finite energy is required to separate them 
infinitely far apart. 

Let us now discuss the effect of turning on a small $J$. 
While the physics within each $xy$ plane remains the same, flipping a $\sigma$ 
spin generates $8$ further defective \emph{vertical} plaquettes. 
Processes that efficiently separated, say, pairs of horizontal defective 
plaquettes in the $J=0$ case, now acquire an additional energy cost that 
scales with the number of flipped spins. 
Indeed, moving defects within a plane leaves behind a wake of flipped 
vertical plaquettes. This ``string'' connecting the defects has an 
energy cost proportional to $J$ times the separation, and the defects become 
{\it linearly confined}. 
A confinement-deconfinement transition occurs when the temperature is large 
enough to overcome the confining energy scale $J$, and the physics of the 
square plaquette model is recovered. 
Therefore, the model with $J\ne 0$ is qualitatively different from $J=0$. 

If we start with $J \ne 0$ and decrease the ratio $J^\prime/J$, we 
expect the same confined physics to survive at sufficiently low temperatures 
(depending on the value of $J$). 
What could perhaps change is the nature of the confinement-deconfinement 
transition at some critical temperature 
$T_c(J,J^\prime)$ 
(indeed, as a function of $J/J^\prime$, we observed using numerical 
simulations what appears to be a change in the transition character from 
second to first order). 
Therefore, we conjecture that the low temperature phase of the anisotropic 
gonihedric model ($J^\prime=0$) discussed in the previous section remains 
essentially unchanged when the ratio $J^\prime/J$ is increased from zero. 
In support of this conjecture, we verified numerically that the location of 
the (putative) first order transition in the isotropic $J^\prime=J$ gonihedric 
model in Ref.~\onlinecite{Johnston2007} is in good agreement with 
the expected transition temperature that is obtained from 2D Ising 
considerations, $T^{\rm iso}_c = 3T_c/2 = 3J/\log(1+\sqrt{2})$ 
(see also Refs.~\onlinecite{Bathas1995},\onlinecite{Koutsoumbas2002}. 
Note that $J = 1/2$ in those references, as well as in 
Ref.~\onlinecite{Johnston2007}). 
%
%

\subsubsection{\label{sec: gaugeable anisotropic gonihedric model}
A (gaugeable) random version of the generalized gonihedric model
              }
Here we discuss yet another generalization of the model in 
Sec.~\ref{sec: J'=0 gonihedric model}. 
That model has an extensive number of minima, namely $2^{L-1}$ for a system
of linear size $L$. The order parameters were $ \langle \s_\ii
\s_{\ii+\hat{\zz}} \rangle = \langle\theta_{\ii+\hat\zz/2}\rangle =
\pm M(T/T_c) $, $\forall\,i_z=1,\ldots,L$ (there was one constraint,
so only $L-1$ of these are independent). Suppose that in 
Eq.~\eqref{eq: generalised gonihedric sigma}
we take $J_{xy}=0$ and generalize the other couplings so that
\begin{eqnarray}
E =&& -\sum_\ii J^{\ii}_{yz}\;
\s_{\ii}\,\s_{\ii+\hat\yy}\,\s_{\ii+\hat\yy+\hat\zz}\,\s_{\ii+\hat\zz}
\nonumber \\ 
&& -\sum_\ii J^{\ii}_{zx}\;
\s_{\ii}\,\s_{\ii+\hat\zz}\,\s_{\ii+\hat\zz+\hat\xx}\,\s_{\ii+\hat\xx}
\nonumber \\ 
=&& -\sum_\bb 
\left(J^{\bb-\hat\zz/2}_{zx}\;
\theta_{\bb}\,\theta_{\bb+\hat\xx}
+
J^{\bb-\hat\zz/2}_{yz}\;
\theta_{\bb}\,\theta_{\bb+\hat\yy}
\right) 
, 
\label{eq: generalised random gonihedric sigma}
\end{eqnarray}
with $J^{\ii}_{zx}=\pm J$ and $J^{\ii}_{yz}=\pm J$ random, but
``gaugeable'', satisfying the condition 
$J^{\ii}_{zx}\,J^{\ii+\hat\xx}_{yz}\,J^{\ii+\hat\yy}_{zx}\,J^{\ii}_{yz}=J^4$. 
In this non-frustrated case the signs of $J^{\bb-\hat\zz/2}_{zx}$ and 
$J^{\bb-\hat\zz/2}_{yz}$ in 
Eq.~\eqref{eq: generalised random gonihedric sigma} can all be gauged out, 
and the thermodynamic behaviour of the system is the same as in 
Sec.~\ref{sec: J'=0 gonihedric model}, undergoing a phase transition 
precisely at the 2D Ising transition temperature $T_c=2J/\log(1+\sqrt{2})$, 
with a $2^{L-1}$ degenerate ``ordered'' phase. However, in this model
it is impossible to write down the order parameter in terms of 
local products of $\sigma$ variables, without using non-local strings
of products of the random variables $J^{\bb-\hat\zz/2}_{zx}=\pm J$ and 
$J^{\bb-\hat\zz/2}_{yz}=\pm J$. This is an example of a gaugeable 
glass.~\cite{footnote:gaugeable_glass,footnote:gaugeable_glass_bis} 
%
%

\subsection{\label{sec: KCM example - variational states}
SMF Hamiltonian and variational excited states
           }
Let us now consider the $J^\prime = 0$ limit of the generalised gonihedric 
model discussed in Sec.~\ref{sec: J'=0 gonihedric model} from the quantum 
mechanical perspective (illustrated in greater detail for the 
Sherrington-Kirkpatrick model 
in App.~\ref{app: two-body examples}). 
[For similar considerations on the isotropic gonihedric model, 
we refer the reader to Appendix~\ref{app: gonihedric model}.] 

Recall that the Ising degrees of freedom $\sigma_i$ live on the sites 
$i$ of a cubic lattice (with periodic boundary conditions, for 
simplicity), and the energy is proportional to the sum over all 
\emph{vertical} plaquette products, that is 
\bea
E 
= 
-J \sum_\bb 
\left( 
  \theta_{\bb}\,\theta_{\bb+\hat\xx} 
  +
  \theta_{\bb}\,\theta_{\bb+\hat\yy} 
\right) 
, 
\eea
where we introduced the Ising variables $\theta_{\bb = \ii +
\hat{\zz}/2} \equiv \s_{i}\,\s_{i+\hat\zz}$, as in Eq.~\eqref{eq:vertical}.

Assuming single spin flip Glauber dynamics, and following the steps outlined 
in App.~\ref{app: two-body examples} (see Ref.~\onlinecite{Castelnovo2005}), 
we arrive at the associated quantum SMF Hamiltonian 
\bea
H_{\rm SMF} 
\!\!&=&\!\! 
\sum_{\ii} 
  \frac{1}{2 \cosh\left[ \frac{\beta \Delta E_\ii}{2} \right]}
    \left\{\vphantom\sum
      \exp\left[ -\frac{\beta \Delta E_\ii}{2} \right] 
      - 
      \sigma^x_\ii 
    \right\}
\nonumber \\ 
\!\!&=&\!\! 
\sum_{\bb} 
  \frac{1}{2 \cosh\left[ \frac{\beta \Delta E_\bb}{2} \right]}
    \left\{\vphantom\sum
      \exp\left[ -\frac{\beta \Delta E_\bb}{2} \right] 
      - 
      \theta^x_\bb \, \theta^x_{\bb-\hat\zz} 
    \right\}
\nonumber \\ 
\Delta E_\bb 
\!\!&=&\!\! 
2J 
\left[\vphantom\sum
  \theta^z_{\bb}\,\theta^z_{\bb+\hat\xx} 
  +
  \theta^z_{\bb}\,\theta^z_{\bb-\hat\xx} 
  +
  \theta^z_{\bb}\,\theta^z_{\bb+\hat\yy} 
  +
  \theta^z_{\bb}\,\theta^z_{\bb-\hat\yy} 
\right. 
\nonumber \\ 
&& 
\left.\vphantom\sum 
  + 
  \theta^z_{\bb-\hat\zz}\,\theta^z_{\bb-\hat\zz+\hat\xx} 
  +
  \theta^z_{\bb-\hat\zz}\,\theta^z_{\bb-\hat\zz-\hat\xx} 
\right. 
\nonumber \\ 
&& 
\left.\vphantom\sum 
  +
  \theta^z_{\bb-\hat\zz}\,\theta^z_{\bb-\hat\zz+\hat\yy} 
  +
  \theta^z_{\bb-\hat\zz}\,\theta^z_{\bb-\hat\zz-\hat\yy} 
\right] 
\equiv 
\Delta E_\ii 
, 
\eea
and the corresponding GS wavefunction is given by the superposition 
\beq
\llvert\psi_0\rrangle 
= 
\sum_\C 
\frac{  
  \exp\left\{
        \frac{\beta J}{2} 
	\sum_\bb 
          \left( 
            \theta^z_{\bb}\,\theta^z_{\bb+\hat\xx} 
            +
            \theta^z_{\bb}\,\theta^z_{\bb+\hat\yy} 
          \right) 
  \right\} 
}{\sqrt{Z}}
  \llvert\C\rrangle 
, 
\eeq
where 
$
Z 
= 
\sum_\C 
\exp\{
  \beta J 
  \sum_\bb 
    ( 
      \theta^z_{\bb}\,\theta^z_{\bb+\hat\xx} 
      +
      \theta^z_{\bb}\,\theta^z_{\bb+\hat\yy} 
    ) 
\}
$. 

As discussed in Sec.~\ref{sec: J'=0 gonihedric model}, the behaviour of the 
ground state of this model is effectively described by decoupled 2D classical 
Ising layers, which is particularly evident in the $\theta$ spin language. 
As a result, there is a critical value $T_c$ where the gap closes in the 
quantum system, and the fidelity susceptibility 
exhibits a singular behavior (namely, 
$\chi_\F \sim C_v^{\rm Ising}$, see Sec.~\ref{sec: fidelity}), 
diverging logarithmically as the transition is approached. 

Below $T_c$, the classical system becomes massively degenerate, with 
energy barriers between the lowest energy states that scale with 
the linear size of the system (see also 
Ref.~\onlinecite{Savvidy2000}). 
These degenerate states give rise to an equal number of slow 
relaxing modes, and therefore to an equal number of low-lying excited states 
in the associated quantum system, whose energy tends to zero in the 
thermodynamic limit. 

Using the variational approach explained in detail in 
Appendix~\ref{app: variational excited states}, we can explicitly find an 
upper bound to a number of low-lying eigenstates of $H_{\rm SMF}$ for 
$T < T_c$ equal to the number of thermodynamic energy minima in the original 
classical system. 
The upper bound tends to zero in the thermodynamic limit, 
thereby proving that the dynamical classical system has many 
relaxing modes whose decay rates vanish in the thermodynamic limit 
(These are equivalent to the degenerate eigenstates discussed in 
Ref.~\onlinecite{Gaveau1998}, where the equality between the numbers of free 
energy minima and of low-lying dynamical states is proven rigorously. 
See also Ref.~\onlinecite{Biroli2001} for a characterisation of 
metastable states in finite size systems in the original classical language.) 

\begin{figure}[ht]
\begin{center}
\includegraphics[width=0.98\columnwidth]
                {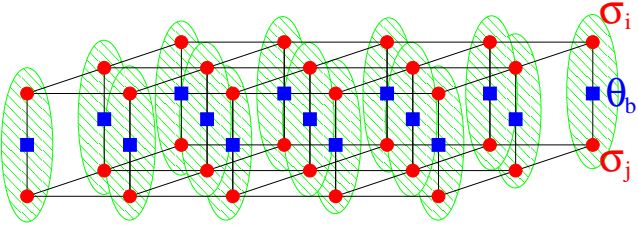}
\end{center}
\caption{
\label{fig: sigma2theta vertical} 
Illustration of the mapping from $\sigma$ to $\theta$ spins. 
}
\end{figure}

The order parameter distinguishing the different 
free energy minima below $T_c$ is the vector $(M_1,\ldots,M_L)$, where 
$M_\ell = \pm \vert M(T/T_c) \vert$ is the average magnetisation in each plane. 
[Recall that a non-vanishing magnetisation of the $\theta$ 
spins along one plane corresponds to a non-vanishing expectation 
value of the sum of products of nearest-neighbour $\sigma$ spin pairs, 
one immediately above and one immediately below that $\theta$ spin plane 
(see Fig.~\ref{fig: sigma2theta vertical})]. 
Following the steps in Appendix~\ref{app: variational excited states}, 
we expect low-lying variational excited states of the form 
\begin{widetext}
\bea
\llvert\psi_{\{n\}}\rrangle 
&=& 
\frac{1}
     {\sqrt{\llangle 
              \llvert
	        O_{\{n\}}(\C) 
	      \rrvert^2 
	    \rrangle_{\rm th} Z}}
\sum_\C 
  O_{\{n\}}(\C) 
  \exp\left\{
        \frac{\beta J}{2} \sum_\bb 
          \left( 
            \theta^z_{\bb}\,\theta^z_{\bb+\hat\xx} 
            +
            \theta^z_{\bb}\,\theta^z_{\bb+\hat\yy} 
          \right) 
  \right\} 
  \llvert\C\rrangle 
\nonumber \\ 
\O_{\{n\}}(\C) 
&=& 
\prod^{L}_{\ell=1}
\left[ 
  \tanh\left( 
    \sum^L_{n,m=1} \theta^z_{n\hat\xx+m\hat\yy+(\ell+1/2)\hat\zz} 
  \right) 
\right]^{n_\ell} 
\qquad\quad 
(n_\ell = 0,1) 
. 
\label{eq: simplest lowest-lying states}
\eea
\end{widetext}
Notice that the argument of the hyperbolic tangent in $\O_{\{n\}}(\C)$ is 
the $z$ component of the magnetisation of the $\theta$ spins in a given 
plane, 
$M_\ell = \sum^L_{n,m=1} \theta^z_{n\hat\xx+m\hat\yy+(\ell+1/2)\hat\zz}$. 
Since $n_\ell = 0,1$, for $\ell=1,\ldots,L$, and given the constraint that 
$\prod^L_{\ell=1} M_\ell$ must be positive, there are $2^{L-1}$ independent 
variational wavefunctions. 

By construction, $\langle \psi_{\{n\}} \vert \psi_0 \rangle = 0$, and 
$\langle \psi_{\{n\}} \vert \psi_{\{n^\prime\}} \rangle = 0$, 
$\forall\,\{n\}\neq\{n^\prime\}$, and eigenstates of $H_{\rm SMF}$ must exist 
with energies 
$
\Delta_{\{n\}} 
\leq 
\langle \psi_{\{n\}} \vert H_{\rm SMF} \vert \psi_{\{n\}} \rangle
$. 
Given that the thermodynamic properties of the anisotropic gonihedric model 
are controlled by the classical 2D Ising model, one can straightforwardly 
follow the steps in Appendix~\ref{app: variational excited states}, to find 
that an upper bound to $\Delta_{\{n\}}$ is given by 
\bea
\Delta_{\{n\}}
&\leq& 
\sum_{\kk} 
\frac{\llangle 
        \vert\O_{\{n\}}(\C)\vert^2 
        -
        \O_{\{n\}}^*(\C) \O_{\{n\}}(\C_\kk) 
      \rrangle_{\rm th}}
     {\llangle \vert\O\vert^2 \rrangle_{\rm th}} 
\nonumber \\ 
&\lesssim& 
\begin{cases}
\O(L^3) & T > T_c 
\\ 
 & 
\\
L^3 \frac{\sum e^{-\vert M(\C) \vert} \, e^{-\beta E_\C}}
     {\sum e^{-\beta E_\C}}
\sim 
L^3\,e^{-\alpha L} & T < T_c 
\end{cases}
\label{eq: gap bound anisotropic gonihedric}
\eea
where the configuration $\C_\kk$ is obtained from the configuration $\C$ upon 
changing the sign of the $\theta$ spins at sites $\kk$ and $\kk+\hat\zz$; 
$\llangle \ldots \rrangle_{\rm th}$ is a classical thermal equilibrium 
average (used here in the mathematical sense, i.e., 
$\llangle M_\ell \rrangle_{\rm th} = 0$ 
even if the system undergoes spontaneous symmetry breaking); 
and $\alpha$ is a positive constant. 
%
%

\section{\label{sec: conclusions} 
Conclusions
        }
In this paper we use a well-known mapping of classical stochastic processes 
onto quantum Hamiltonians~\cite{symmetric_W_refs,RK_like_refs} 
(see Ref.~\onlinecite{Castelnovo2005} for a detailed constructive approach) 
to argue that dynamical glass transitions can be interpreted in the 
quantum mechanical language as static zero-temperature phase 
transitions where a number (exponentially large in system size) 
of excited states collapse onto the ground state. The quantum
mechanical perspective allows one to accurately define what a 
{\it dynamical} glass transition means, and could provide new avenues to 
understand the consequences of the transition on the {\it static} 
properties of the system. Here we proposed to use concepts from 
quantum information, such as entanglement entropy and fidelity 
susceptibility, as tools to help uncover ``hidden'' order in glasses, 
which may not be accessible through local order parameters. 

In mapping the stochastic dynamics of classical systems to quantum 
Hamiltonians, the relaxation spectrum of the classical system corresponds to 
the excitation spectrum of the quantum model. 
Any interesting (i.e., non-exponentially decaying) dynamical behaviour of the 
former requires necessarily the vanishing of one or more relaxation rates. 
That is, the appearance of modes, other than the equilibrium 
distribution, which fail to relax during the stochastic process. 
Understanding such dynamical features in the classical system is 
tantamount to understanding the behaviour of the spectrum in 
the associated zero-temperature, static quantum system.~\cite{footnote:d+1} 

In the past few years a great deal of knowledge has been accumulated in the 
study of quantum phase transitions. In particular, it has become clear that 
there are transitions for which it is not possible to define local order 
parameters. Nevertheless, there are ways of detecting and 
characterizing such transitions without resorting to an order parameter. 
One recent tool is the concept of fidelity, which is constructed from the
overlap of the ground state wavefunction at two infinitesimally close values 
of the coupling constant that drives the transition. Here we showed that the 
fidelity susceptibility, in quantum systems derived via the mapping 
from classical Forker-Planck evolutions, is precisely the heat capacity of 
the classical system. Therefore, we argue that, if the fidelity picks up 
quantum transitions in local Hamiltonians, then a dynamical transition in 
a classical system with local energies and local dynamics must be 
accompanied by a thermodynamic signature (singularity) in the heat
capacity. We note that there are examples of disordered 
systems where dynamic and thermodynamic transitions differ, such 
as the fully connected $p$-spin glass model;~\cite{Montanari2006} 
however, these are non-local systems. 
While there is no rigorous proof that all quantum phase 
transitions give rise to singularities in the fidelity, we know of no 
counter example in local Hamiltonians, including quantum disordered 
and topologically ordered systems. Indeed one can view our results as 
forcing a marriage between quantum information and non-equilibrium 
glassy physics, with strong implications from one field into the 
other. One is thus pressed to accept that either 
A) there exist strange undetectable quantum phase transitions in local 
Hamiltonians without local order parameters or even fidelity singularities, 
or B) there is a true thermodynamic glass transition, without 
necessarily any local order parameters, but with a heat capacity 
singularity nonetheless. 

Another tool that has proven effective to probe exotic states 
without \textit{a priori} knowledge of an order parameter is the 
von Neumann entanglement entropy. 
Here we studied its behavior 
in quantum states constructed via the mapping from stochastic dynamics. 
We showed that, in the case of glassy phases with
an extensive number of non-relaxing modes (in the thermodynamic limit), 
the entanglement entropy of a bipartition of the system 
obeys the (conventional) area law both above
and below the transition, but with a prefactor that jumps across the
phase transition. In addition, scaling properties of the von Neumann
entropy as a function of the size of the subsystems 
can be used to reveal a correlation length that diverges at the 
dynamical phase transition. 

Finally, in this paper we introduced an example of a non-disordered spin 
model that exemplifies well the concepts arising from the classical to 
quantum mapping, and it exhibits several of the features typical of a 
``glassy'' spectral collapse. 
The model, discussed in Sec.~\ref{sec: KCM example}, is an 
example of a system with purely local interactions that can undergo a 
thermodynamic transition at finite temperature into a phase with a 
(sub)extensive number of equilibrium states. 
The model has $2^{L-1}$ degenerate equilibrium states, and its 
(gaugeable) disordered version is an exactly 
solvable ideal spin glass, with purely local interactions and a 
(sub)extensive number of statistically equivalent (from a local point of view) 
equilibrium states. 

Understanding exotic zero-temperature quantum 
phase transitions is not obviously simpler than studying glass 
transitions. The mapping discussed in this paper is not a magic wand, but 
rather a change in perspective. Yet sometimes a change in perspective is 
what one needs to gain new insight on a long standing problem. 


%
%

\section*{Acknowledgments}
We would like to thank Malcolm Kennett and Benoit Dou\c{c}ot for their 
encouraging enthusiastic reaction at the early stages of this project. 
We are indebted to Leticia Cugliandolo, Jorge Kurchan, Giulio Biroli, 
Robert Jack and Juan P. Garrahan for several insightful discussions and 
constructive criticisms. 
This work was supported in part by EPSRC 
Grant No.~GR/R83712/01 and by EPSRC Postdoctoral Research Fellowship 
EP/G049394/1 (C.~Castelnovo), and in part by DOE Grant 
DEFG02-06ER46316 (C.~Chamon). C.~Castelnovo acknowledges the 
hospitality of the Condensed Matter Theory Visitor's Program at Boston 
University. 
%
%
\appendix

\section{\label{app: collapsing states} 
Nature of the collapsing states
        }
As we discussed earlier, any dynamical transition
requires a collapse of relaxation rates, i.e., of 
eigenenergies in the associated quantum Hamiltonian $H_{\rm SMF}$. 
Let us focus here on the nature of these collapsing states, in particular 
their relation to spontaneous symmetry breaking, broken ergodicity, and 
glassiness. 

Consider first the conventional case of a classical thermodynamic 
symmetry breaking transition with a local order parameter. 
Below the transition temperature, the classical free energy landscape 
develops distinct minima separated by energy barriers $\sim\!L^\alpha$, 
for some exponent $\alpha\le d$. The number $\N$ of such minima is controlled 
by the broken symmetry, and it is typically finite (i.e., it does not grow 
with the size of the system): 
if the broken symmetry is discrete, there are a finite number of such states; 
if the symmetry is continuous, the manifold of degenerate states is labelled 
by a continuous variable. 
This in turn leads to $\N$ low lying eigenstates of $H_{\rm SMF}$, 
with energies that scale as $\varepsilon_1 \sim e^{-aL^\alpha}$ 
(recall that the ground state energy is by construction $\varepsilon_0=0$). 
Take for instance the Ising ferromagnet in 
Appendix~\ref{app: variational excited states}. 

The nature of the spectrum above these exponentially degenerate
states depends on the dynamics governing the relaxation within a 
minimum. For instance, coarsening leads to a power law
$\varepsilon_{\rm ex} \sim\!L^{-z}$ level spacing. Notice that one can
argue for Goldstone modes ($\varepsilon_{\rm ex} \sim\!L^{-b}$) in the
continuous symmetry breaking case, but one can still have
algebraic (in $1/L$) energy levels even in discrete systems,
as the $H_{\rm SMF}$ spectrum depends on the dynamics 
(consider for instance diffusive modes in presence of conserved quantities). 

The exponentially large times required to switch between broken symmetry 
minima give rise to broken ergodicity in the thermodynamic limit. 
However, the presence of such diverging time scales is immaterial, because 
they can be observed only in a fine tuned system. 
The presence of a local order parameter means that an infinitesimal 
local perturbation~\cite{footnote:local_perturbation} to the energy of the 
classical system produces a finite separation between the energies of the 
different minima -- with consequent removal of the diverging time scales. 
In studying the behaviour of the system, we are thus justified to pick one 
of the symmetry-broken states over the others. 

Such a relation between symmetry breaking order parameters and 
broken ergodicity is well known. There are however a number of other 
scenarios that lead to a collapse of relaxation rates in classical 
Markov processes. 
\vspace{0.3cm}

(i) \textbf{Systems with broken symmetry where the degeneracy scales
with system size} -- 
This is the case for the example without quenched disorder discussed in 
Sec.~\ref{sec: KCM example}. 
An important difference between this case and the one where the
degeneracy does not scale with system size is that it is not possible
to split all the degeneracy by applying an infinitesimal local perturbation. 
However, it is still usually possible to select a unique GS, and one could 
argue that the physics in the end is not much different from a 
conventional spontaneous symmetry breaking transition. 
To illustrate this, let us consider the example in 
Sec.~\ref{sec: J'=0 gonihedric model}, where the system behaves as decoupled 
2D Ising ferromagnetic layers. 
An infinitesimal uniform magnetic field $h$ lowers the energy of the uniformly 
magnetised state by an amount $\gtrsim h L^2$ with respect to all other 
states that are degenerate for $h=0$. 
Note that, crucially, $h L^2$ diverges in the thermodynamic limit for any 
infinitesimal but \emph{fixed} value of $h$. 

(ii) \textbf{Disordered glassy systems} -- 
If we consider well known examples of glassy systems such as the 
Sherrington-Kirkpatrick model, discussed in App.~\ref{app: two-body examples}, 
a degeneracy emerges at low temperatures, at least if we adopt Parisi's 
picture, which scales with system size. 
The minima are statistically equivalent from a local perspective, and 
unlike the case of magnetic ordering in an Ising ferromagnet, there is no 
local operator that we can apply uniformly across the system to lift the 
degeneracy. 
Clearly if one knew the lowest energy spin configuration in one of the 
minima, a magnetic field could be tailored to favour that precise 
configuration energetically. Whilst this is indeed a perturbation given by 
the sum of local operators, the values of the field are random but fixed 
specifically for each and every spin in the lattice. 
Such random field contains highly non-local information, and 
we feel that this should not be called a local 
perturbation.~\cite{footnote:local_perturbation} 

(ii) \textbf{Systems with gaugable (i.e., not frustrating) disorder} -- 
The leading difference between glass transitions and symmetry breaking 
transitions -- namely, the ability to select a unique GS by means of 
infinitesimal local perturbations -- can be removed by introducing disorder 
in a gaugeable fashion (see for example the model in 
Sec.~\ref{sec: gaugeable anisotropic gonihedric model}). 
A gaugeable disorder does not spoil the thermodynamic behaviour, 
but without full knowledge of the gauge transformation, one can no longer 
distinguish the degenerate states below $T_c$ using local perturbations. 
For a model such as the anisotropic gonihedric one, with a GS degeneracy 
that scales with the size of the system, the addition of gaugeable disorder 
yields a model that becomes indeed very similar to a disordered glassy 
system. One might then wonder whether the gaugeability of the disorder 
does in fact constitute a material difference between the models, or if 
the physics is ultimately the same from a qualitative point of view. 

(iv) \textbf{Systems with topological degeneracy} -- 
For completeness, it is interesting to compare the scenarios above 
(where the collapsing spectrum was drawn from classical stochastic processes) 
with yet a different case drawn from quantum mechanics, 
although there is no \textit{a priori} connection with classical stochastic 
processes in this case. 
Consider a zero temperature quantum system which undergoes a phase transition 
into a topologically ordered phase. In this phase, the system develops a 
topological degeneracy which is ultimately non-local. Not only does any 
infinitesimal local perturbation fail to produce a finite lifting of the 
degeneracy, but actually any local operator must have identical expectation 
values across all the GSs, and vanishing matrix elements between any 
two degenerate states. 
For example, the reader can think of modifications of the 
toric code Hamiltonian.~\cite{Kitaev2003} As shown in 
Ref.~\onlinecite{Castelnovo2008a}, one can indeed construct SMF-like 
quantum Hamiltonians that exhibit topological order in regions of their 
phase diagrams. 
In systems where the number of topologically degenerate states scales with 
the size of the system (e.g., in models similar to those in 
Sec.~\ref{sec: KCM example}, briefly outlined in 
Appendix~\ref{app: gonihedric model}), the spectral collapse at a 
transition into the topological phase is qualitatively that of a perfect 
glass; so perfect that it yields no measurable consequences! 
Indeed, the classical free energy minima are \emph{exactly} identical 
from a statistical point of view, and the diverging time scales for the 
system to go from one minimum to another are 
\emph{immaterial to all locally measurable quantities} 
(e.g., the Edwards-Anderson order parameter vanishes on both sides of the 
transition). 
In order to see a diverging time scale, one would need, say, to compute 
non-local (loop-loop) autocorrelation functions. 

To conclude, when the collapse of the energy spectrum 
is due to the spontaneous symmetry breaking, or to the 
emergence of topological order, the system goes through a quantum phase
transition in $H_{\rm SMF}$ into a phase with a manifold of 
exponentially degenerate states $\Delta E\sim e^{-aL^\alpha}$. 
In this paper, we argue that also a dynamical (glass) transition 
becomes a well-defined static phase transition in the associated quantum 
mechanical language. 
%
%

\section{\label{app: two-body examples}
The Sherrington-Kirkpartick model, revisited
        }
Let us illustrate the concepts discussed in Sec.~\ref{sec: method} and 
in Sec.~\ref{sec: q_EA} with an established example, 
the Sherrington-Kirkpartick model.~\cite{Sherrington1975} 
Consider a system of $N$ Ising spins $\sigma_i$, $i=1,\ldots,N$, with phase 
space $\Omega_N = \{+1,-1\}^N$, $\vert \Omega_N \vert = 2^N$, 
subject to two-body interaction terms with coupling constants 
$\{ J_{ij} \}_{i,j=1,\ldots,N}$, $J_{ii} = 0$, $\forall\,i$: 
\beq
E(\{ \sigma \}) = - \sum_{i,j} J_{ij} \sigma_i \sigma_j 
, 
\label{eq: classical 2body energy} 
\eeq
where each $J_{ij}$ ($= J_{ji}$) is independently Gaussian distributed with 
zero mean and standard deviation $\sim 1/N$. 
Let us also assume that the dynamic processes are limited to single spin 
flip events, governed by Glauber dynamics. That is, a transition between 
two configurations $\C,\Cp\in\Omega_N$ differing by a single spin flip 
occurs with probability: 
\beq
P_{\C \to \Cp} 
= 
\frac{e^{-\beta (E_\Cp -E_\C)/2}}
     {2 \cosh[\beta (E_\Cp -E_\C)/2]} 
. 
\label{eq: Glauber dynamics}
\eeq
%
%


The SMF quantum Hamiltonian is then given by~\cite{Castelnovo2005} 
\bea
\hat{H}_{\rm SMF} 
&=& 
\sum_{(\C,\Cp)} 
\frac{1}
     {2 \cosh[\beta (E_\Cp -E_\C)/2]} 
\nonumber \\ 
&\times&
\left\{ 
  e^{-\beta (E_\Cp -E_\C)/2} \llvert\C\rrangle\llangle\C\rrvert 
  - 
  \llvert\C\rrangle\llangle\Cp\rrvert 
\right\} 
, 
\label{eq: SMF Hamilonian gen}
\eea
in the $\{ \llvert\C\rrangle \}$ basis, where $(\C,\Cp)$ stands for pairs of 
configurations connected by a single spin flip, 
$\llvert\Cp\rrangle = \hat{\sigma}^x_i \llvert\C\rrangle$, $\exists i$. 

Using Eq.~\eqref{eq: classical 2body energy}, 
and introducing the local fields $h_i = \sum_j J_{ij} \sigma_j$, 
we can write 
\bea
E_\Cp - E_\C 
&=& 
2 \sum_j J_{ij} \sigma_i \sigma_j 
= 
2 h_i \sigma_i 
\nonumber \\ 
\cosh[\beta (E_\Cp -E_\C)/2] 
&=& 
\cosh[\beta h_i] 
, 
\nonumber 
\eea
and Eq.~\eqref{eq: SMF Hamilonian gen} becomes 
\bea
\hat{H}_{\rm SMF} 
&=& 
\sum_{i,\C} 
\frac{1}
     {2 \cosh[\beta h_i]} 
\left\{\vphantom{\sum} 
  e^{-\beta h_i \sigma_i} \llvert\C\rrangle\llangle\C\rrvert 
  - 
  \llvert\C\rrangle\llangle\C\rrvert \hat{\sigma}^x_i 
\right\} 
. 
\nonumber
\eea

Up to now, the variables $\sigma_i$ and $h_i$ are c-numbers. 
Introducing the corresponding operators $\hat{\sigma}^z_i$ and 
$\hat{h}_i = \sum_j J_{ij} \hat{\sigma}^z_j$, 
one can simplify the notation and do away with the summation over $\C$ by 
observing that 
\bea
\sum_\C \frac{1}{2 \cosh[\beta h_i]} 
e^{-\beta h_i \sigma_i} \llvert\C\rrangle\llangle\C\rrvert 
&=& 
\frac{e^{-\beta \hat{h}_i \hat{\sigma}^z_i}}{2 \cosh[\beta \hat{h}_i]}
\\ 
\sum_\C \frac{1}{2 \cosh[\beta h_i]} 
\llvert\C\rrangle\llangle\C\rrvert \hat{\sigma}^x_i 
&=& 
\frac{\hat{\sigma}^x_i}{2 \cosh[\beta \hat{h}_i]}
, 
\eea
where we used the fact that $[\hat{h}_i,\hat{\sigma}^x_i]=0$ 
(recall that $J_{ii}$ = 0). 
We finally arrive at the expression, 
\bea
\hat{H}_{\rm SMF} 
&=& 
\sum_{i} 
\frac{1}
     {2 \cosh[\beta \hat{h}_i]} 
\left\{ 
  e^{-\beta \hat{h}_i \hat{\sigma}^z_i} 
  - 
  \hat{\sigma}^x_i 
\right\} 
. 
\label{eq: 2body SMF hamiltonian}
\eea

Similarly, the GS wavefunction can be written 
as~\cite{Castelnovo2005} 
\bea
\llvert\psi_0\rrangle 
&=& 
\frac{1}{\sqrt{Z}} \sum_\C e^{-\beta E_\C/2} \llvert\C\rrangle 
\nonumber \\ 
&=& 
\sum_\C 
  \frac{1}{\sqrt{Z}} 
    \exp\left\{\frac{\beta}{2} \sum_i \hat{h}_i \hat{\sigma}^z_i \right\} 
    \llvert\C\rrangle 
, 
\label{eq: GS wavefunction}
\eea
where $Z = \sum_\C \exp\{\beta \sum_i \hat{h}_i \hat{\sigma}^z_i \}$. 

{}For convenience of notation, we will drop the $\hat{\cdot}$ symbols from 
now onward, and all $h_i$, $\sigma^z_i$, and $\sigma^x_i$ will be understood 
as quantum mechanical operators. 

Given that the two operators $e^{-\beta h_i \sigma^z_i}$ and $\sigma^x_i$ in 
Eq.~\eqref{eq: 2body SMF hamiltonian} do not commute, obtaining the 
spectrum of the system for all values of the parameter $\beta$ is a tall 
order, in spite of knowing the exact GS wavefunction. 
In the following, we will investigate analytically the low and high 
temperature limits ($\beta \ll 1$, and $\beta \gg 1$).
%
%

\subsection{\label{sec: high-temperature limit}
High-temperature behaviour
           }
The limit of $\beta \ll 1$ can be studied by expanding the 
Hamiltonian to first order in $\beta$: 
\bea
H_{\rm SMF} 
\simeq 
\frac{N}{2} 
- 
\frac{\beta}{2}
\sum_{i} h_i \sigma^z_i 
- 
\frac{1}{2}
\sum_{i} \sigma^x_i 
+ \O(\beta^2) 
. 
\label{eq: 2body SMF hamiltonian -- beta << 1}
\eea
Up to corrections of order $\beta$, the GS is given by the (unique) 
eigenvector of $\sigma^x_i$ with eigenvalues $+1$ for all $i$, 
and the gap above it is equal to $1$. 
According to Eq.~\eqref{eq: corr bound} and Eq.~\eqref{eq: EA parameter}, 
this scenario corresponds to exponentially fast decaying autocorrelation 
functions, and vanishing Edwards-Anderson order parameter, as expected in the 
high-temperature phase of the dynamical classical system. 

[Note that the discussion so far applies in general to any classical Ising 
model with two-body interactions and Glauber dynamics.] 
%
%

\subsection{\label{sec: low-temperature limit}
Low-temperature behaviour
           }
The limit of $\beta \gg 1$ is more subtle, and it is convenient to approach 
it by looking at the GS wavefunction. 
We shall make use of working assumptions inspired by Parisi's work in 
Ref.~\onlinecite{Parisi1983}: at low enough temperatures, the classical phase 
space effectively divides into ``basins of influence'' of the different minima 
$a=1,\ldots,\N$ of the free energy. 
That is, one can divide the partition function of the system $Z=\sum_a Z_a$, 
where all the relevant states for a given minimum $a$ fall into the 
corresponding partial partition function $Z_a$, and the ambiguity in assigning 
all the states in between minima is immaterial as their total weight in the 
partition function is negligible. 
For example, the reader could have in mind a 2D Ising ferromagnet, 
where below $T_c$ one can safely divide the partition function into a 
positive and a negative magnetisation contribution 
(see for instance Sec.3.1 in Ref.~\onlinecite{Kurchan2009}). 

Let us define the probability of being in a given basin, 
$P_a = Z_a / Z$ \ ($\sum_a P_a = 1$). 
The GS wavefunction of the quantum SMF Hamiltonian can then be written as 
\bea
\llvert\psi_0\rrangle 
&\approx& 
\sum_a \sqrt{P_a} \: \llvert\phi_a\rrangle 
\nonumber \\ 
\llvert\phi_a\rrangle 
&\approx& 
\frac{1}{\sqrt{Z_a}} \sum_{\C \in a} e^{-\beta E_\C/2} \llvert\C\rrangle 
. 
\label{eq: GS wavefunction low-T}
\eea

Notice that the partial wavefunctions $\llvert\phi_a\rrangle$ have the same 
amplitudes as the GS wavefunction, but they involve only the relevant states 
in basin $a$. 
The assumption that the partition function can be divided into basins implies 
that local changes to a configuration $\C$ belonging to one basin will either 
produce a configuration in the same basin, or a configuration $\Cp$ whose 
weight is negligible, $\exp[-\beta E_\Cp / 2]/Z_a \ll e^{-N}$. 
Since the off-diagonal term in the SMF Hamiltonian (with single spin-flip 
dynamics) is local, one can verify that this assumption results in 
$\llvert\phi_a\rrangle$ being an approximate eigenvector of $H_{\rm SMF}$ with 
vanishing eigenvalue. 

The fact that no configuration state $\llvert\C\rrangle$ appears in more 
than one $Z_a$ guarantees that the $\llvert\phi_a\rrangle$ are orthogonal to 
each other. 
They form therefore a basis of the low-lying manifold of eigenstates of the 
quantum system. 

We can then compute the Edwards-Anderson order parameter within this 
assumption [see Eq.~\eqref{eq: EA parameter}]: 
\bea
q_{\rm EA}(\O) 
&=& 
\sum^{\N-1}_{n=1} \: 
\left\vert 
  \llangle\psi_0\rrvert \hat{\O} \llvert\psi_n\rrangle 
\right\vert^2 
\nonumber \\ 
&=& 
\sum^{\N-1}_{n=0} \: 
\left\vert 
  \llangle\psi_0\rrvert \hat{\O} \llvert\psi_n\rrangle 
\right\vert^2 
- 
\left\vert 
  \llangle\psi_0\rrvert \hat{\O} \llvert\psi_0\rrangle 
\right\vert^2 
\nonumber \\ 
&=& 
\llangle\psi_0\rrvert \hat{\O} 
  \left(
    \sum^{\N-1}_{n=0} \: \llvert\psi_n\rrangle \llangle\psi_n\rrvert 
  \right)
\hat{\O} \llvert\psi_0\rrangle 
- 
\left\vert 
  \llangle\psi_0\rrvert \hat{\O} \llvert\psi_0\rrangle 
\right\vert^2 
\nonumber \\ 
&=& 
\llangle\psi_0\rrvert \hat{\O} 
  \hat{\mathcal{P}}
\hat{\O} \llvert\psi_0\rrangle 
- 
\left\vert 
  \llangle\psi_0\rrvert \hat{\O} \llvert\psi_0\rrangle 
\right\vert^2 
, 
\eea
where the operator $\hat{\mathcal{P}}$ is nothing but a projector onto the 
manifold of the $\N$ low-lying states. We can therefore write it as 
$\hat{\mathcal{P}} = \sum^\N_{a=1} \llvert\phi_a\rrangle\llangle\phi_a\rrvert$. 
Substituting into the previous equation, we arrive at the result 
\bea
q_{\rm EA}(\O) 
&\approx& 
\llangle\psi_0\rrvert \hat{\O} 
  \sum^\N_{a=1} \llvert\phi_a\rrangle\llangle\phi_a\rrvert
\hat{\O} \llvert\psi_0\rrangle 
- 
\left\vert 
  \llangle\psi_0\rrvert \hat{\O} \llvert\psi_0\rrangle 
\right\vert^2 
\nonumber \\ 
&\approx& 
\sum^\N_{a=1} 
  P_a \llvert \llangle\phi_a\rrvert\hat{\O}\llvert\phi_a\rrangle \rrvert^2 
- 
\left\vert 
  \sum^\N_{a=1} 
    P_a \llangle\phi_a\rrvert\hat{\O}\llvert\phi_a\rrangle 
\right\vert^2 
, 
\nonumber 
\eea
where we used the fact that the operators $\hat{\O}$ are necessarily diagonal 
in the configuration basis $\{\llvert\C\rrangle\}$ (since they derive from 
classical observables $\O$), and therefore 
$\llangle\phi_a\rrvert\hat{\O}\llvert\phi_b\rrangle = 0$, $\forall\,a \neq b$. 

Notice that the expectation value 
$\llangle\phi_a\rrvert \hat{\O} \llvert\phi_a\rrangle$ corresponds to the 
ensemble average of the classical observable $\O$ in the $a$th basin of the 
free energy, 
$\llangle\O\rrangle_a = \sum_{\C \in a} \O_\C \exp[-\beta E_\C] / Z_a$. 
Therefore, 
\bea
q_{\rm EA}(\O) 
&\approx& 
\sum^\N_{a=1} 
  P_a \llangle\O\rrangle^2_a  
- 
\left\vert 
  \sum^\N_{a=1} 
    P_a \llangle\O\rrangle_a 
\right\vert^2 
. 
\eea
If we consider for example $\O=\sigma^z_i$, the second term vanishes in the 
absence of an explicit $\mathbb{Z}_2$ symmetry breaking term. 
By taking the average over all sites, 
\bea
q_{\rm EA}(\O) 
&\approx& 
\sum^\N_{a=1} P_a 
  \sum^N_{i=1} \frac{\llangle\sigma^z_i\rrangle^2_a}{N} 
\equiv 
\sum^\N_{a=1} P_a \, 
  q^a_{\rm EA}  
, 
\eea
we recover indeed the Edwards-Anderson order parameter 
in Parisi's formulation,~\cite{Parisi1983} averaged over all basins. 
%
%

\subsection{\label{sec: griffiths singularities}
Griffiths singularities
           }
As noted in Sec.~\ref{sec: q_EA}, 
a finite value of $q_{\rm EA}$ implies that the local
static susceptibility diverges in the low-temperature phase.
However, this is a sufficient but not necessary condition.
For example, Griffiths singularities -- characteristic of quantum 
disordered systems -- can appear above the transition, and cause the 
static susceptibility to diverge while $q_{\rm EA}$ 
remains zero.~\cite{Griffiths_refs} 

Although investigating these issues is beyond the scope of the present paper, 
it is interesting to take a brief look at the specific example of the SK 
model, in the limit of $\beta \ll 1$, 
Eq.~\eqref{eq: 2body SMF hamiltonian -- beta << 1}: 
\bea
H_{\rm SMF}
\sim
-
\frac{\beta}{2}
\sum_{i,j} J_{ij} \sigma^z_j \sigma^z_i
-
\frac{1}{2}
\sum_{i} \sigma^x_i
.
\label{eq: SK expanded}
\eea
It is precisely the Hamiltonian of a quantum Ising spin glass in a 
transverse field, with mean 
field random interactions $J_{i,j}$. 
If it is the case that the small-$\beta$ approximation captures the physics 
of the full Hamiltonian $H_{\rm SMF}$ in the high-temperature phase 
$\beta < \beta_c$, one would not expect any Griffiths singularities appearing 
in the disordered phase of this model.  
Indeed, in Eq.~\eqref{eq: SK expanded} the local static 
susceptibility diverges only at the transition.~\cite{Rieger1997} 

Although they may not play a role in the SK model, Griffiths
singularities are likely to appear in other SMF Hamiltonians associated with
disordered dynamical systems (e.g., systems where $J_{ij}$ is short 
ranged).~\cite{Garnerone2009} 
In such cases, it will be interesting to understand whether the Griffiths
singularities actually correspond to observable features in the original
classical system, possibly relating to the presence of separate
characteristic temperatures, as is the case for the dynamical, the 
thermodynamic one-step replica-symmetry-breaking and the 
full replica-symmetry-breaking transitions in the $p \geq 3$ Ising spin 
glass model. 
%
%

\section{\label{app: variational excited states} 
Variational approach to the slowest relaxing modes
        }
Here we show how one can construct variationally the collapsing states 
(slowest relaxing modes) in the associated quantum system. 
We shall consider for simplicity the case of a classical nearest-neighbour 
Ising model, in which case the ground state in the quantum system is only 
two fold degenerate below $T_c$ 
(see also Refs.~\onlinecite{Gaveau1998},\onlinecite{Biroli2001} and 
references therein). 
The results we obtain are however more general, and in 
Sec.~\ref{sec: KCM example - variational states} we show how they can be used 
to find a subextensive set of collapsing state in an ideal glass system. 

The SMF Hamiltonian for the nearest-neighbour Ising model is a special 
example of the class of Hamiltonians discussed in 
App.~\ref{app: two-body examples}, namely where 
$h_i = J \sum_{\llangle ij \rrangle} \sigma^z_j$, $J$ being the coupling 
constant in the classical system. 
Eq.~\eqref{eq: 2body SMF hamiltonian} reduces then to 
\bea
H_{\rm SMF} 
&=& 
\sum_{i} 
\frac{
  \left\{ 
    e^{-\beta J \sum_{\llangle ij \rrangle} \sigma^z_j \sigma^z_i} 
    - 
    \sigma^x_i 
  \right\} 
     }
     {2 \cosh[\beta J \sum_{\llangle ij \rrangle} \sigma^z_j]} 
, 
\label{eq: 2d Ising SMF hamiltonian}
\eea
and the relative GS wavefunction is 
\bea
\llvert\psi_0\rrangle 
&=& 
\sum_\C 
  \frac{1}{\sqrt{Z}} 
    \exp\left\{
      \frac{\beta J}{2} \sum_{\llangle ij \rrangle} \sigma^z_j \sigma^z_i 
    \right\} 
    \llvert\C\rrangle 
, 
\eea
where 
$Z=\sum_\C \exp\{\beta J \sum_{\llangle ij \rrangle} \sigma^z_j \sigma^z_i \}$. 
[Remark on the notation: the sums over $\langle ij\rangle$ are over all
nearest neighbors pairs of sites $i,j$, unless there is an outside sum
over the site $i$, namely $\sum_i \sum_{\langle ij\rangle}$; 
in which case $\sum_{\langle ij\rangle}$ stands for
the sum over all $j$s that are neighbors to the $i$ site in the external
sum $\sum_i$.]

Recall that, by construction, the GS $\llvert\psi_0\rrangle$ is annihilated by 
each term in curly brackets in Eq.~\eqref{eq: 2d Ising SMF hamiltonian}. 
The existence of other state(s) $\llvert\psi_1\rrangle$ asymptotically 
degenerate with $\llvert\psi_0\rrangle$ 
(in the thermodynamic limit) means that 
%
%
\begin{subequations}
\bea
\llangle \psi_0 \vert \psi_1 \rrangle 
&=&
0 
\label{eq: excited state conditions 1}
\\ 
H_{\rm SMF} \llvert \psi_1 \rrangle
&=& 
\Delta_1 \llvert \psi_1 \rrangle 
\label{eq: excited state conditions 2}
\\
\lim_{N \to \infty} \Delta_1 &=& 0
, 
\label{eq: excited state conditions 3}
\eea
\label{eq: excited state conditions}
\end{subequations}
where $N$ is the size of the system. 

While finding an exact expression for $\llvert \psi_1 \rrangle$ is rather 
difficult, one can attempt a variational approach. First of all, let us 
write $\llvert \psi_1 \rrangle$ as 
\bea
\llvert\psi_1\rrangle 
= 
\sum_\C 
  \frac{\O(\C)}{\sqrt{\llangle \vert\O\vert^2 \rrangle_{\rm th} Z}} 
    \exp\left\{
      \frac{\beta J}{2} \sum_{\llangle ij \rrangle} \sigma^z_j \sigma^z_i 
    \right\} 
    \llvert\C\rrangle 
, 
\label{eq: excited  wavefunction}
\eea
for a generic set of coefficients $\O(\C)$. 
The notation $\llangle \ldots \rrangle_{\rm th}$ stands for a thermal average 
in the original classical system, namely 
\bea
\llangle \vert\O\vert^2 \rrangle_{\rm th} 
&=& 
\frac{1}{Z} 
\sum_\C 
  \vert\O(\C)\vert^2
  \exp\{\beta J \sum_{\llangle ij \rrangle} \sigma^z_j \sigma^z_i \} 
. 
\eea
Notice that this thermal average is to be interpreted here as a mathematical 
summation over \emph{all} spin configurations $\C$, irrespective, for instance, 
of the loss of ergodicity that occurs under spontaneous symmetry breaking 
(i.e., $\llangle \sigma^z_i \rrangle_{\rm th} = 0$ both above and below $T_c$). 

One can then show that $\llangle \psi_0 \vert \psi_1 \rrangle$ is
proportional to the thermal average $\llangle \O \rrangle_{\rm
th}$. So long as we find a set of coefficients $\O(\C)$ where
$\llangle \O \rrangle_{\rm th} \to 0$ for $N \to \infty$, we satisfy
the first condition in Eq.~\eqref{eq: excited state conditions} in the
thermodynamic limit. That is, the condition of orthogonality in
~\eqref{eq: excited state conditions} requires finding an observable
$\O$ whose average over all configurations vanishes in the
thermodynamic limit. (Notice that if $\llangle \O \rrangle_{\rm th}=0$
because of a symmetry, as is the case in the Ising model, then
$\llangle \psi_0 \vert \psi_1 \rrangle=0$ even for finite $N$.)

Given $\llangle \O \rrangle_{\rm th} = 0$, one can immediately find an upper 
bound for the energy gap above the GS of the system, 
\bea
\Delta_1 
\leq 
\llangle \psi_1 \rrvert H_{\rm SMF} \llvert \psi_1 \rrangle 
. 
\eea

Clearly the process can be iterated. Given a second observable $\O^\prime$ 
that satisfies $\llangle \O^\prime \rrangle_{\rm th} = 0$ and 
$\llangle \O \O^\prime \rrangle_{\rm th} = 0$, the state 
$\llvert\psi_2\rrangle$ constructed from $\O^\prime$ in the same way that 
$\llvert\psi_1\rrangle$ was constructed from $\O$ is orthogonal to both 
$\llvert\psi_0\rrangle$ and $\llvert\psi_1\rrangle$, and another eigenstate 
must exist which differs in energy from the GS by 
$\Delta_2 \leq \llangle \psi_2 \rrvert H_{\rm SMF} \llvert \psi_2 \rrangle$

Let us look in more detail at the structure of these upper bounds. 
Using the Hamiltonian in Eq.~\eqref{eq: 2d Ising SMF hamiltonian}, we obtain 
\begin{widetext}
\bea
\llangle \psi_1 \rrvert H_{\rm SMF} \llvert \psi_1 \rrangle 
&=& 
\frac{1}{\llangle \vert\O\vert^2 \rrangle_{\rm th} Z} 
\sum_{\C,\Cp} 
  \O^*(\C)
    \exp\left\{
      \frac{\beta J}{2} \sum_{\llangle ij \rrangle} \sigma^z_j \sigma^z_i 
    \right\} 
  \O(\C^\prime)
    \exp\left\{
      \frac{\beta J}{2} \sum_{\llangle ij \rrangle} \sigma'^z_j \sigma'^z_i 
    \right\} 
\nonumber \\ 
&\times& 
\llangle\C\rrvert 
\sum_{k} 
\frac{
  \left\{ 
    e^{-\beta J \sum_{\llangle kl \rrangle} \sigma^z_l \sigma^z_k} 
    - 
    \sigma^x_k 
  \right\} 
     }
     {2 \cosh[\beta J \sum_{\llangle kl \rrangle} \sigma^z_l]} 
\llvert\Cp\rrangle 
\\ 
&=& 
\frac{1}{\llangle \vert\O\vert^2 \rrangle_{\rm th} Z} 
\sum_{k} 
\sum_{\C} 
  \left[\vphantom\sum
    \vert\O(\C)\vert^2 
    -
    \O^*(\C) \O(\C_k)
  \right] 
  \frac{
    \exp\left\{
      \beta J \sum^{i,j \neq k}_{\llangle ij \rrangle} \sigma^z_j \sigma^z_i 
    \right\} 
       }
       {2 \cosh[\beta J \sum_{\llangle kl \rrangle} \sigma^z_l]} 
\\ 
&=& 
\frac{1}{\llangle \vert\O\vert^2 \rrangle_{\rm th}} 
\sum_{k} 
\frac{1}{Z} 
\sum_{\C} 
  \frac{
    \left[\vphantom\sum
      \vert\O(\C)\vert^2 
      -
      \O^*(\C) \O(\C_k)
    \right] 
    \exp\left\{
      -\beta J \sum_{\llangle kl \rrangle} \sigma^z_l \sigma^z_k 
    \right\} 
       }
       {2\cosh[\beta J \sum_{\llangle kl \rrangle} \sigma^z_l]} 
  \exp\left\{
    \beta J \sum_{\llangle ij \rrangle} \sigma^z_j \sigma^z_i 
  \right\} 
\nonumber \\ 
&=& 
\sum_{k} 
\frac{\llangle 
        \frac{
          \left[\vphantom\sum
            \vert\O(\C)\vert^2 
            -
            \O^*(\C) \O(\C_k)
          \right] 
          \exp\left\{
            -\beta J \sum_{\llangle kl \rrangle} \sigma^z_l \sigma^z_k 
          \right\} 
             }
             {2\cosh[\beta J \sum_{\llangle kl \rrangle} \sigma^z_l]} 
      \rrangle_{\rm th}}
     {\llangle \vert\O\vert^2 \rrangle_{\rm th}} 
, 
\label{eq: gap bound 1}
\eea
\end{widetext}
where $\C_k$ denotes the spin configuration that obtains from $\C$ upon 
changing the sign of $\sigma^z_k$. 
[Recall that, in presence of an overall summation over $k$, the sum 
$\sum_{\llangle kl \rrangle}$ above stands for a summation over all sites 
$l$ that are nearest-neighbours to $k$.] 

Since the ratio between the exponential and the hyperbolic cosine in 
Eq.~\eqref{eq: gap bound 1} is of the form $x(x+1/x)^{-1}$ with $x$ positive, 
irrespective of the sign of $\sigma^z_k$, we can simplify the upper bound 
\bea
\Delta_1 
&\leq& 
\llangle \psi_1 \rrvert H_{\rm SMF} \llvert \psi_1 \rrangle 
\nonumber \\ 
&\leq& 
\sum_{k} 
\frac{\llangle 
        \vert\O(\C)\vert^2 
        -
        \O^*(\C) \O(\C_k) 
      \rrangle_{\rm th}}
     {\llangle \vert\O\vert^2 \rrangle_{\rm th}} 
. 
\label{eq: gap bound 3}
\eea
Therefore, if we are interested in the lowest lying eigenstates, we need to 
choose $\O$ such that 
$
\sum_k \llangle 
        \vert\O(\C)\vert^2 
        -
        \O^*(\C) \O(\C_k) 
      \rrangle_{\rm th} 
\ll 
\llangle \vert\O\vert^2 \rrangle_{\rm th}
$. 

Specifically for systems that undergo a symmerty breaking phase transition, 
such as the classical $d$-dimensional Ising model 
with nearest-neighbour interactions ($d > 1$), 
a convenient choice for $\O(\C)$ that satisfies the condition 
$\llangle\O\rrangle_{\rm th} = 0$ is an odd function of the order parameter, 
such as the magnetisation of the system, $M(\C) \equiv \sum_i \sigma^z_i$ 
(recall that $\llangle\ldots\rrangle_{\rm th}$ stands for the mathematical 
ensemble average summed over \emph{all} spin configurations $\C$). 

In the case of the Ising model, a convenient choice to obtain a low 
energy bound is $\O(\C) = \tanh[M(\C)]$. 
Note that $M(\C)$ is the extensive magnetisation of the sample, and 
$\tanh^2[M(\C)] \geq \tanh^2(1)$ so long as $M(\C) \neq 0$. 
The denominator in Eq.~\eqref{eq: gap bound 3} is bounded from below by 
\bea
\llangle \vert\O\vert^2 \rrangle_{\rm th} 
&=& 
\llangle \tanh^2[M(\C)] \rrangle_{\rm th} 
\nonumber \\ 
&\geq&
\tanh^2(1) 
\frac{\sum_{\{\C \, : \, M(\C) \neq 0\}} e^{-\beta E_\C}}
     {\sum e^{-\beta E_\C}}
\nonumber \\ 
&\gtrsim&
\tanh^2(1) 
, 
\eea
where the ratio between the partition functions of the Ising model with and 
without the constraint $M(\C) \neq 0$ tends to $1$ in systems with an 
even number of sites, and it is identically $1$ in systems with an odd number 
of sites. 

We can thus focus only on the numerator in Eq.~\eqref{eq: gap bound 3}, 
\bea
&&\!\!\!\!\!\! 
\vert\O(\C)\vert^2 
-
\O^*(\C) \O(\C_k)
= 
\nonumber \\ 
&&\!\!\!\!\!\! 
\qquad 
= 
\tanh^2[M(\C)] 
-
\tanh[M(\C)] \tanh[M(\C_k)] 
\nonumber \\ 
&&\!\!\!\!\!\! 
\qquad 
= 
\tanh^2[M(\C)]
\left\{ 
  1 
  -
  \frac{\tanh[M(\C_k)]}{\tanh[M(\C)]} 
\right\} 
. 
\eea
While in general this quantity can be negative, we know by construction that 
the GS energy of the SMF Hamiltonian is exactly zero, and therefore any 
upper bound to $\Delta_1$ must be non-negative. 
Under single spin flip dynamics, $M(\C_k) = M(\C) \pm 2$. 
Assuming $M(\C) > 2$ and $M(\C_k) < M(\C)$ for convenience, 
\bea
&&\!\!\!\!\!\! 
\vert\O(\C)\vert^2 
-
\O^*(\C) \O(\C_k)
\leq 
\nonumber \\ 
&&\!\!\!\!\!\! 
\qquad 
\leq 
1 
-
\frac{\tanh[M(\C_k)]}{\tanh[M(\C)]} 
\nonumber \\ 
&&\!\!\!\!\!\! 
\qquad 
\leq 
1 
-
\tanh[M(\C_k)] 
\nonumber \\ 
&&\!\!\!\!\!\! 
\qquad 
\leq 
2 e^{-2 M(\C_k)} 
. 
\eea
Similarly for $M(\C) > 2$ and $M(\C_k) > M(\C)$, one obtains 
\bea
\vert\O(\C)\vert^2 
-
\O^*(\C) \O(\C_k)
\leq 
2 e^{-2 M(\C)} 
, 
\eea
and analogous results hold for the two corresponding cases with $M(\C) < -2$, 
with opposite sign in the exponent (i.e., $e^{2 M(\C)}$ and $e^{2 M(\C_k)}$, 
respectively). 

Therefore, the thermal average appearing in the numerator of 
Eq.~\eqref{eq: gap bound 3} 
can be interpreted as the ratio between the partition functions of the 
classical Ising model with and without an additional energy term that 
penalises magnetised states, 
\beq
\llangle 
  \vert\O(\C)\vert^2 
  -
  \O^*(\C) \O(\C_k) 
\rrangle_{\rm th} 
\;\lesssim\; 
\frac{\sum e^{-\vert M(\C) \vert} \, e^{-\beta E_\C}}
     {\sum e^{-\beta E_\C}}
. 
\nonumber
\eeq
While this ratio is finite in the paramagnetic phase, for $T < T_c$ it
becomes exponentially small in the system size $\sim\exp(- \gamma
L^{d-1})$, with $\gamma$ a non-universal constant.  This exponential
penalty is the least one can incur [as opposed to a cost depending on
the extensive magnetization and thus $~\sim\exp(- m L^{d})$], and it
is the cost of a domain wall that keeps $M(\C)\sim 0$ in the
$e^{-\vert M(\C) \vert}$ term. So the bound on the gap $\Delta_1$ from
this variational state is:
\beq
\Delta_{1} 
\;\lesssim\; 
\begin{cases}
\O(L^d) & T > T_c 
\\ 
 & 
\\
L^d\,e^{-\gamma L^{d-1}} & T < T_c 
\end{cases}
. 
\eeq
Notice that the bound for $T > T_c$ is rather loose, and we could have
found a much better bound of ${\cal O}(1)$ by other means. However, it
is the bound for the ordered phase $T < T_c$ that is of interest
here. The spectrum collapses exponentially in the system size for $d>1$.

The corresponding eigenvector for the Ising model is 
\bea
\llvert\psi_1\rrangle 
= 
\sum_\C 
  \frac{\tanh[\sum_i\sigma^z_i]}
       {\sqrt{\llangle \tanh^2[\sum_i\sigma^z_i] \rrangle_{\rm th}}} 
  \frac{
    \exp\left\{
      \frac{\beta J}{2} \sum_{\llangle ij \rrangle} \sigma^z_j \sigma^z_i 
    \right\} 
       }
       {\sqrt{Z}}
    \llvert\C\rrangle 
. 
\nonumber 
\eea
Qualitatively, one can understand this result by noting that 
$
\tanh[\sum_i\sigma^z_i] 
/ 
\sqrt{\llangle \tanh^2[\sum_i\sigma^z_i] \rrangle_{\rm th}} 
\sim 
\pm 1
$, 
and the excited state $\llvert\psi_1\rrangle$ is essentially the 
antisymmetric superposition of the positive and negative magnetisation 
valleys in the free energy. 
As expected, below $T_c$ the two valleys become ergodically disconnected, 
and it takes an exponential time in the size for the system to migrate 
from one valley to the other, leading to two distinct slow-relaxing modes, 
and therefore two lowest-lying states in the associated SMF Hamiltonian. 
For any finite size system, the actual GS is given by the symmetric 
(nodeless) superposition of the two states, whereas the antisymmetric 
superposition lies at a slightly higher energy $\Delta_1$. 
However, we expect (and indeed we just showed) that $\Delta_1 \to 0$ 
in the thermodynamic limit $N \to \infty$, for $T < T_c$. 

We illustrated this constructive approach to a variational low-lying excited 
state in the simple case of a nearest neighbour Ising model. Given 
some \textit{a priori} knowledge on the order parameter that distinguishes the 
free energy minima at low temperatures, it can be straightforwardly applied 
to find upper bounds to the lowest energy levels, say, in quantum Hamiltonians 
derived from classical systems that undergo a dynamical transition. 
In the case of the Sherrington-Kirkpatrick model, briefly discussed in 
App.~\ref{app: two-body examples}, this can be done only at a formal level, 
since we do not know explicitly the actual form of the non-local operators that select 
one valley over another. On the other hand, in 
Sec.~\ref{sec: KCM example} we illustrate how the lowest energy states can be 
explicitly constructed, say, in kinetically constrained models 
without disorder. 

[The reader might be interested in comparing our derivation of the 
excited states and of the upper bound to the spectral gap using a 
variational principle in quantum mechanics with the more elaborate but 
exact calculation of the relaxation spectrum using transition currents 
in Sec.3.2.1 in Ref.~\onlinecite{Kurchan2009}.] 
%
%

\section{\label{app: gonihedric model} 
Quantum views on the gonihedric model
        }
In Sec.~\ref{sec: KCM example} we introduced the generalised gonihedric model, 
and we discussed a few specific cases. 
Here we present two distinct dual descriptions of the isotropic limit, 
$J_{xy}=J_{yz}=J_{zx}=J$, and we show how the relative SMF Hamiltonians relate 
to toric-code-like models in three dimensions (i.e., infinitely massive 
$\mathbb{Z}_2$ quantum gauge theories). 
%
%

\subsection{\label{app: sigma gonihedric model} 
Gonihedric model
           }
Let us first recall the structure of the generalised gonihedric model. 
The Ising degrees of freedom $\sigma_i = \pm 1$ live on the sites of a cubic 
lattice, where we shall label the square plaquettes by $p$. 
The energy of the system can then be written as 
\beq
E 
= 
- J \sum_{p} \prod_{i \in p} \sigma_i 
, 
\label{eq: gonihedric sigma energy app}
\eeq
where $i \in p$ label the four sites at the corners of $p$. 
Notice that flipping a plane of $\s$ spins does not change the energy in 
Eq.~\eqref{eq: gonihedric sigma energy app}. This symmetry of the model 
results in a minimal degeneracy of isoenergetic configurations that 
scales as $2^{3L-2}$. 

Assuming Glauber single spin flip dynamics, a few algebraic 
steps~\cite{Castelnovo2005} as in the example discussed in 
Sec.~\ref{sec: KCM example} lead to the associated SMF Hamiltonian 
\bea
H^{(0)}_{\rm SMF} 
&=& 
\sum_{i} 
  \frac{1}{2 \cosh\left[ \frac{\beta \Delta E_i}{2} \right]}
    \left\{\vphantom\sum
      \exp\left[ -\frac{\beta \Delta E_i}{2} \right] 
      - 
      \sigma^x_i 
    \right\}
\nonumber \\ 
\Delta E_i 
&=& 
2 J \sum_{\{ p \,:\, i \in p \}} 
    \prod_{j \in p} \sigma^z_j 
. 
\eea
%
%

\subsection{\label{app: theta gonihedric model} 
Gonihedric model in the bond-dual spin language
           }
\begin{figure}[ht]
\begin{center}
\includegraphics[width=0.6\columnwidth]
                {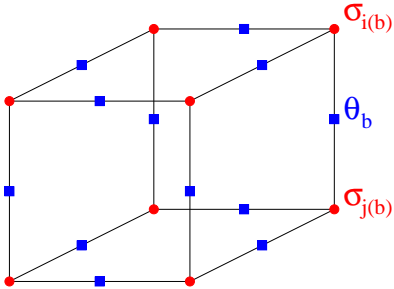}
\end{center}
\caption{
\label{fig: sigma2theta} 
Illustration of the mapping from $\sigma$ to $\theta$ spins. 
}
\end{figure}

An alternative description of the same classical system can be formulated 
in terms of Ising spins living on the bonds $b$ of the lattice, 
$\theta_b = \sigma_{i_-(b)} \sigma_{i_+(b)}$, where $i_\pm(b)$ are the two 
sites adjacent to bond $b$, as illustrated in Fig.~\ref{fig: sigma2theta}. 
In Sec.~\ref{sec: J'=0 gonihedric model}, discussing the anisotropic 
gonihedric model, we used only the $\theta$ spins on the vertical bonds of 
the cubic lattice. Indeed, with the addition of a horizontal plane 
of $\sigma$ spins, they are sufficient to describe the entire system. 
Here we define instead a $\theta$ spin on \emph{each} bond of the lattice. 
Whilst the notation becomes highly redundant, it demonstrates an 
interesting analogy between the SMF Hamiltonian for the isotropic gonihedric 
model and a class of topologically ordered systems called toric 
code models.~\cite{Kitaev2003} 

In this language, given a plaquette with corners $i_1,\ldots,i_4$, and bonds 
$b_1,\ldots,b_4$, the corresponding 4-body energy contribution 
$\sigma_{i_1}\sigma_{i_2}\sigma_{i_3}\sigma_{i_4}$ becomes a 2-body term 
that can be written as $\theta_{b_1}\theta_{b_3}$ or 
$\theta_{b_2}\theta_{b_4}$. 
However, not all $\{\theta_b\}$ configurations are allowed by the mapping. 
Indeed, for every plaquette we have the gauge constraint 
$\theta_{b_1}\theta_{b_2}\theta_{b_3}\theta_{b_4} = 1$. 
In addition, the product of all the $\theta$ spins along a winding line on 
the lattice (periodic boundary conditions are assumed) must be positive. 
Even if there are $3N$ $\theta$ spins for a system of $N$ $\sigma$ spins 
(i.e., $N$ sites of the cubic lattice), the constraints reduce the number of 
independent $\theta$s, and ultimately one can see that the mapping is 
$1$-to-$2$, since one can obtain the value of any $\sigma$ spin given 
all the $\theta$s plus \emph{one} reference $\sigma$. 
The energy of the system can be written as 
\beq
E 
= 
- \frac{J}{2} \sum_{[[ b b^\prime ]]} \theta_b \theta_{b^\prime} 
, 
\eeq
where the notation $[[ b b^\prime ]]$ stands for two parallel bonds $b$ and 
$b^\prime $ belonging to the same plaquette. 

Note that Glauber single spin flip dynamics in the $\sigma$ spins translates 
into a `star' flipping operation, that changes the sign of all $6$ $\theta_b$ 
spins adjacent to a common site $i$ ($b \in i$). 
The associated SMF Hamiltonian assumes the form 
\bea
H^{(1)}_{\rm SMF} 
&=& 
\sum_{i} 
  \frac{1}{2 \cosh\left[ \frac{\beta \Delta E_i}{2} \right]}
    \left\{\vphantom\sum
      \exp\left[ -\frac{\beta \Delta E_i}{2} \right] 
      - 
      \prod_{b \in i} \theta^x_b 
    \right\}
\nonumber \\ 
\Delta E_i 
&=& 
J \sum_{b \in i} \sum_{[[ b b^\prime]]} \theta^z_b \theta^z_{b^\prime} 
. 
\eea
Notice that all terms in the Hamiltonian commute with the constraints, namely 
the product of all $\theta^z$ components of the spins around a plaquette 
$p$ is the identity, $\prod_{b \in p} \theta^z_b = \openone$, and that the 
product of $\theta^z$ along any winding loop $\gamma$ also gives the 
identity, $\prod_{b \in \gamma} \theta^z_b = \openone$. 
We can therefore extend the zero-temperature SMF quantum system to the 
unconstrained Hilbert space generated by the spin-$1/2$ $\theta$ degrees of 
freedom, provided that we add an appropriately large energy cost to states 
that violate the constraints: 
\bea
H^{(1)}_{\rm SMF} 
&=& 
\sum_{i} 
  \frac{1}{2 \cosh\left[ \frac{\beta \Delta E_i}{2} \right]}
    \left\{\vphantom\sum
      \exp\left[ -\frac{\beta \Delta E_i}{2} \right] 
      - 
      \prod_{b \in i} \theta^x_b 
    \right\}
\nonumber \\ 
&-& 
\sum_p \prod_{b \in p} \theta^z_b 
- 
\prod_{b \in \gamma_x} \theta^z_b
-
\prod_{b \in \gamma_y} \theta^z_b
-
\prod_{b \in \gamma_z} \theta^z_b
, 
\label{eq: theta SMF}
\eea
where $\gamma_x$, $\gamma_y$ and $\gamma_z$ are three generic winding loops 
along the $x$, $y$ and $z$ direction, respectively. 

Is is interesting to notice that the limit $J=0$ in Eq.~\eqref{eq: theta SMF} 
is nothing but Kitaev's toric code 
Hamiltonian in 3D,~\cite{Kitaev2003,Hamma2005b,Castelnovo2008b} 
with the addition of explicit non-local operators (the products along the 
$\gamma_\alpha$ winding loops) that select one out of the $8$ topological 
sectors. 
The (gapped) topological phase corresponds to the high-temperature limit of 
the classical system, and it survives for small values of $\beta$ up to 
a phase transition (the transition in the classical gonihedric 
model~\cite{gonihedric_first_order}) where topological order 
is lost, and yet \emph{neither of the two neighbouring phases appears to be 
captured by a local order parameter}. 
%

\subsection{\label{app: S gonihedric model} 
Gonihedric model in the face-dual spin language
           }
\begin{figure}[ht]
\begin{center}
\includegraphics[width=0.8\columnwidth]
                {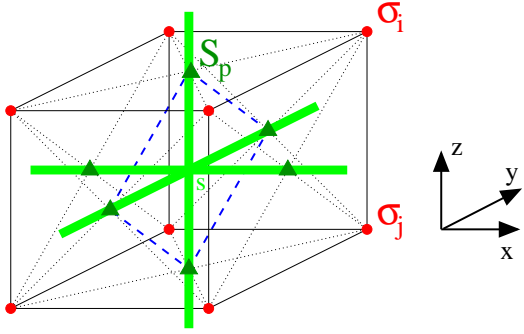}
\end{center}
\caption{
\label{fig: sigma2S} 
Illustration of the mapping from $\sigma$ to $S$ spins. 
The $S$ spins living at the plaquette centres in the original lattice can be 
also seen as bond degrees of freedom in the dual cubic lattice (thick 
light lines in the figure) formed by the centres $s$ of the unit cubic 
cells in the original lattice. 
The four $S$ spins belonging to a planar star on the dual lattice, 
centred at $s$ and perpendicular to the $x$ direction, are highlighted by 
the dashed blue lines. 
}
\end{figure}

Finally, there is another useful description of the isotropic gonihedric 
model, formulated in terms of Ising spins living on the plaquettes $p$ of the 
lattice, $S_p = \prod_{i \in p} \sigma_{i}$, as illustrated in 
Fig.~\ref{fig: sigma2S}. 
Note that the $S$ spins live on the bonds of the cubic lattice formed by the 
centres of the unit cubic cells in the original lattice. 
In this language, the energy becomes a trivial sum of 1-body terms 
\beq
E 
= 
- J \sum_{p} S_p 
, 
\eeq

\begin{figure}[ht]
\begin{center}
\includegraphics[width=0.85\columnwidth]{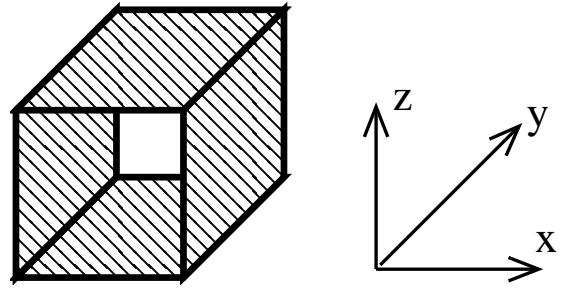}
\end{center}
\caption{
\label{fig: annular identities} 
Illustration of the local constraints in the dual $S_p$ spin languange 
in the $y$ direction: 
the product of the $S_p$ spins on the shaded plaquettes must always be equal 
to $1$. 
Similarly for the $x$ and $z$ directions. 
}
\end{figure}

As above, not all $\{S_p\}$ configurations are allowed by the mapping. 
For every unit cubic cell in the lattice there are three constraints that 
must be satisfied. Namely, the product of the four plaquette spins on the 
faces of the cell parallel to a lattice direction must be equal to the 
identity, for each direction $x$, $y$, and $z$, as illustrated in 
Fig. ~\ref{fig: annular identities}
%
%
(see Ref.~\onlinecite{Savvidy1996} for a related duality transformation). 
In the dual cubic lattice, these identities correspond to the condition 
$\prod_{\{p \in s \,:\, p \perp \alpha\}} S_p = +1$, where $s$ denotes a 
generic dual lattice site, $p \in s$ labels the $6$ plaquette sites on the 
adjacent bonds, and $p \perp \alpha$ means that the dual bond on which $p$ 
lives is perpendicular to the lattice direction $\alpha=x,y,z$. 
In addition, the product of all plaquette spins $S_p$ along any strip 
(equivalent to a spin ladder, in the original $\sigma$ spins) winding around 
the system is also equal to $1$. 
In the end, one can see that the mapping is 
$1$-to-$(3L-2)$, since we can obtain the value of any $\sigma$ spin given 
all the $S$ spins plus all the $\sigma$ spins along \emph{three} reference 
straight lines winding around the system, and crossing all at one point in 
the lattice. 
Note that $3L-2$ is precisely the degeneracy of the lowest energy 
states in the gonihedric model. Indeed, changing the sign of all 
the spins belonging to the same lattice plane is an exact symmetry of the 
classical energy, and it allows one to construct $3L-2$ distinct lowest energy 
configurations starting from, say, the fully magnetised one 
$\{ \sigma_i = +1 \}$. All such configurations correspond to the same 
fully magnetised $\{ S_p = +1 \}$ plaquette spin configuration, in agreement 
with the $1$-to-$(3L-2)$ nature of the mapping. 

The Glauber single spin flip dynamics in the $\sigma$ spins translates into 
a flipping operation that changes the sign of all the $S_p$ spins 
on plaquettes that have the flipped spin $\sigma_i$ at one of their corners 
($\{ p \,:\, i \in p \}$). 
Alternatively, this can be seen as the operation that flips all $S_p$ living 
on the $12$ edges of a cubic unit cell in the dual lattice (thick light green 
lines in Fig.~\ref{fig: sigma2S}), centred at $i$. 
The associated SMF Hamiltonian assumes thus the form 
\bea
H^{(2)}_{\rm SMF} 
\!&=&\! 
\sum_{i} 
  \frac{1}{2 \cosh\left[ \frac{\beta \Delta E_i}{2} \right]}
    \left\{\vphantom\sum
      \exp\left[ -\frac{\beta \Delta E_i}{2} \right] 
      - 
      \!\!\!\!\!\prod_{\{ p \,:\, i \in p \}}\!\!\! S^x_p 
    \right\}
\nonumber \\ 
\Delta E_i 
\!&=&\! 
2J \sum_{\{ p \,:\, i \in p \}} S^z_p 
. 
\eea

All terms in the Hamiltonian commute with the constraints, namely 
the products of the $S^z$ components of the spins around planar stars 
and straight winding lines $\gamma$ (perpendicular to the centres of 
plaquettes in the original lattice) being the identity. 
We can therefore extend the zero-temperature SMF quantum system to the 
unconstrained Hilbert space generated by the spin-$1/2$ $S$ degrees of 
freedom, provided that we add an appropriately large energy cost to states 
that violate the constraints: 
\bea
H^{(2)}_{\rm SMF} 
\!&=&\! 
\sum_{i} 
  \frac{1}{2 \cosh\left[ \frac{\beta \Delta E_i}{2} \right]}
    \left\{\vphantom\sum
      \exp\left[ -\frac{\beta \Delta E_i}{2} \right] 
      - 
      \!\!\!\!\!\prod_{\{ p \,:\, i \in p \}}\!\!\! S^x_p 
    \right\}
\nonumber \\ 
&-& 
\sum_{\alpha = x,y,z} \sum_s \prod_{\{p \in s \,:\, p \perp \alpha\}} S^z_p 
- 
\sum_\gamma 
  \prod_{p \in \gamma} S^z_p 
, 
\label{eq: S SMF}
\eea
where $s$ labels the sites in the dual cubic lattice. 

In the limit of $J=0$, Eq.~\eqref{eq: S SMF} reduces to a 
$\mathbb{Z}_2$ lattice gauge theory different from the one obtained in 
the previous section. 
This gauge theory has a peculiar topological
degeneracy that scales with the size of the system to the power $2/3$
(namely, the sectors are identified by the eigenvalues of the $\sim
3L^2$ $\gamma$ winding loop operators). As in the case of the
$\theta$ spin mapping, Eq.~\eqref{eq: S SMF} contains non-local terms
that select a unique topological sector. The (gapped) topological
phase corresponds to the high-temperature limit of the classical
system, and it survives for small values of $\beta$. 

Now, the $\beta\to\infty$ limit can be recognized as the trivial fully
magnetised state with $S_p=+1$ for all $p$, which is unique and
non-degenerate (in the $S_p$ variables of this face-dual
description). We can distinguish between the two phases using 
the topological entropy defined in 
Refs.~\onlinecite{Levin2006},\onlinecite{Kitaev2006}. 
Here we find $S_{\rm topo}(\beta=0)\ne 0$ and 
$S_{\rm topo}(\beta\to \infty)= 0$. We can 
therefore prove the existence of a phase transition separating these
two phases at some critical temperature $\beta_c$, in support of the 
current evidence based on numerical simulations and cluster mean field 
arguments.~\cite{gonihedric_first_order} 
(See also Ref.~\onlinecite{Pietig1996_97} for an alternative proof of 
the transition.) 
%
%


\begin{thebibliography}{99}
\bibitem{footnote:local_perturbation} 
For the sake of clarity, what we mean here by ``local perturbation'' is an 
energy term proportional to the sum of operators with finite support, which 
differ only by translations. For instance, 
$\sum_i \s_i$ 
or 
$\sum_i \s_i \s_{i+\hat{x}} \s_{i+\hat{y}} \s_{i+\hat{x}+\hat{y}}$. 
According to this definition, a random field does not qualify as a local 
perturbation.
\end{thebibliography}
\end{document}